\documentclass[useAMS, usenatbib]{mn2e}
\usepackage{amsmath}
\usepackage{amssymb}
\usepackage{graphicx}  
\usepackage{pdflscape}
\usepackage{color}
\DeclareMathOperator{\sech}{sech}   
\def\msun{\,M_\odot}
\def\pc{\,{\rm pc}}
\def\kpc{\,{\rm kpc}}
\def\kms{\,{\rm km\,s^{-1}}}
\def\Myr{\,{\rm Myr}}\def\Gyr{\,{\rm Gyr}}
\def\e{{\rm e}}\def\rd{{\rm d}}
\def\feh{\left[{\rm Fe/H}\right]}
\def\afe{\left[\alpha/{\rm Fe}\right]}

\title[Quiescent phase of galactic disc growth]
{The quiescent phase of galactic disc growth}
\author[M. Aumer et al.]
{Michael Aumer \thanks{E-mail:Michael.Aumer@physics.ox.ac.uk (MA)}, James Binney and Ralph Sch\"onrich\\
Rudolf Peierls Centre for Theoretical Physics, 1 Keble Road, Oxford, OX1 3NP, UK}

\begin{document}

\date{Accepted 2016 March 31. Received 2016 March 30; in original form 2016 February 3}

\pagerange{\pageref{firstpage}--\pageref{lastpage}} \pubyear{2016}

\maketitle

\label{firstpage}

\begin{abstract}

We perform a series of controlled $N$-body simulations of growing disc
galaxies within non-growing, live dark matter haloes of varying mass and
concentration. Our initial conditions include either a low-mass disc or a
compact bulge. New stellar particles are continuously added on near-circular
orbits to the existing disc, so spiral structure is continuously excited. To
study the effect of combined spiral and giant molecular cloud (GMC) heating
on the discs we introduce massive, short-lived particles that sample a GMC
mass function. An isothermal gas component is introduced for a subset of the
models. We perform a resolution study and vary parameters governing the GMC
population, the histories of star formation and radial scale growth.  Models
with GMCs and standard values for the disc mass and halo density provide the
right level of self-gravity to explain the age velocity dispersion relation
of the Solar neighbourhood (Snhd). GMC heating generates remarkably
exponential vertical profiles with scaleheights that are radially constant
and agree with observations of galactic thin discs. GMCs are also capable of
significantly delaying bar formation. The amount of spiral induced radial
migration agrees with what is required for the metallicity distribution of
the Snhd. However, in our standard models the outward migrating populations
are not hot enough vertically to create thick discs. Thick discs can form in
models with high baryon fractions, but the corresponding bars are too long,
the young stellar populations too hot and the discs flare considerably.

\end{abstract}


\section{Introduction}

The vertical profile of our Galaxy in the Solar Neighbourhood (Snhd) can be
closely fitted by the sum of two exponentials with scaleheights of $h_{z,{\rm
thin}}\sim 300\pc$ and $h_{z,{\rm thick}}\sim 900\pc$ \citep{gilmore, juric}.
\citet{juric} found that both thin and thick components could be
characterised by exponential radial profiles, with the thick component having
a larger scalelength than the thin component.  Many external disc galaxies
also show vertical profiles that can be decomposed into thin and thick
components \citep{yoachim}, with the thick component having a radial
scalelength that is similar to, or slightly larger than, the scalelength of
its thin counterpart. 

In the solar cylinder the thick component contributes $\sim 30$ per cent of
the local stellar mass surface density, and its stars tend to have larger
random velocities, higher ages, lower metallicities and higher
$\alpha$-element abundances \citep{bensby}. In a plot of $\afe$ versus
$\feh$, two sequences can be identified: a `normal' or `low-$\alpha$'
sequence, which starts around $\feh \sim -0.6$ with a small slope, and a
`high-$\alpha$' sequence, which is nearly flat at low $\feh$ and then bends
more steeply downwards to join the low-$\alpha$ sequence around
$\feh\sim0.2$. \citet{bovyrix2} used data from the SEGUE survey to argue that
the vertical velocity dispersion of stars varies continuously with chemistry,
and that signs of a thin/thick disc dichotomy are an artefact of selection
functions.  However, \citet{nidever14} and \citet{hayden15} found well
defined low and high $\alpha$ abundance sequences in the $(\feh,\afe)$ plane
of the APOGEE survey. While it is tempting to infer two distinct populations
from a bimodality in $\afe$, \citet{sb09a, sb09b} (hereafter SB09a,b) have
shown that such a bimodality is a natural consequence of chemical evolution
timescales. 

The high-$\alpha$ sequence dominates larger altitudes in the disc, but its
importance diminishes towards larger radii. This observational finding has
lead to the conclusion that the $\alpha$-rich component of the Milky Way (MW) has a
comparably shorter scale length \citep{BensbyF, Cheng, bovyrix}. This fading
of the thick high-$\alpha$ component seems to contradict the long photometric
thick disc scale length found by \citet{juric} and the lack of evidence of a
change in the vertical profile as a function of radius both in our Galaxy and
in external disc galaxies \citep{vdkruit, comeron}.  One possibility is that
low $\alpha$ populations flare with increasing radius and take the role of
the thick component in the outer disc (see e.g. \citealp{minchev15}).

The vertical scaleheight of a stellar population in a given galactic
potential is determined by its vertical velocity dispersion.  If we group
nearby stars by age, the velocity dispersion of the group increases with age
(\citealp{stromberg, parenago, wielen}). This phenomenon has been
inferred also for external galactic discs (\citealp{beasley} for M33,
\citealp{dorman} for M31). However, for Snhd stars this increase appears to be
continuous with age and does not indicate a thin-thick bimodality
\citep{holmberg, ab09}.

Currently stars are born very close to the plane with small velocity
dispersions, but over time fluctuations in the Galaxy's gravitational field
cause stars to diffuse from the near-circular orbits of their birth to more
eccentric and inclined orbits, with the consequence that the velocity
dispersion of each coeval group of stars steadily increases. It is possible
that \emph{all} disc stars were born on nearly circular orbits, but it is
also possible that at early times the young Galaxy's gravitational field
fluctuated so strongly that even dense gas could not settle to circular
orbits, so stars had significant random velocities at birth. A major goal of
this paper is to investigate the possible genesis of thick-disc stars
assuming no significant external perturbation throughout the era in which 
nearly all disc stars formed.

Fluctuations in the gravitational field that accelerate stars can arise in
several ways. \citet{spitzer} suggested that an important source of
fluctuations is giant gas clouds.  The subsequent discovery of Giant
molecular clouds (GMCs) with masses $M_{\rm GMC}\sim 10^{5-7}\msun$
vindicated this suggestion. \citet{barbanis} showed that spiral structure
can be a significant source of fluctuating gravitational field, provided it
has either a very high density contrast or is of transient and recurring
nature. Stars can also be heated by merging satellite galaxies 
\citep{gerhardF, toth, velazquez}.

Another key ingredient in the evolution of disc galaxies is bars. In today's
universe around 70 per cent of all disc galaxies, including the MW
\citep{jjb91},  have stellar bars at their centres
\citep{eskridge}.  Bars have been shown to contribute to the radial and
vertical heating in disc galaxies \citep{saha} and to provide a mechanism for
the radial migration of stars \citep{friedli, minchev10}. Bars are believed to be a
natural outcome of the evolution of dynamically cold galactic discs (e.g.
\citealp{miller}).  However, observations of galaxies at redshifts between 0
and 1 have revealed that the fraction of bars was significantly lower several
Gyr ago \citep{sheth}.

Although it is generally believed that a combination of heating processes
mentioned above can explain the thin disc age velocity 
dispersion relation (AVR), the origin of the thick disc
is still heavily debated. Two key questions are: 1) Did thick-disc stars form
in a thin disc which was subsequently heated, or did they form with large
random velocities in a turbulent gas disc \citep{bournaud,forbes}? 2) If they formed
in a thin disc is an external perturber required to endow them with their
present random velocities, or could these have arisen through internal
secular processes, such as scattering off the bar \citep{minchev10} and GMCs
followed by radial migration \citep{sellwoodb} of stars from the hot
inner disc to the cooler outskirts (SB09b)?

In this paper, we seek to understand whether double exponential vertical
profiles of disc galaxies can arise purely from secular evolution, i.e. from
heating by GMCs and by structures such as bars and spirals, which inevitably
occur in a growing disc galaxy \citep{selcar, carsel}. As spiral heating is
expected to contribute little to the vertical heating (see \citealp{martinez}
for a recent confirmation with modern isolated disc simulations) our focus
here is on GMCs, bars and the outward transport of hot populations through
radial migration.  

To study these heating processes, usually idealised, isolated models have
been used. Unfortunately, the majority of these models do not at the same
time include all the ingredients, which have been implicated in disc heating:
growing discs with multiple coeval populations, recurring spiral structure
with evolving properties, a bar, radial migration, and GMCs that at early
times each contain a bigger fraction of the total disc mass than they do
today.

Cosmological, hydrodynamical simulations of the formation of disc galaxies
do, in principle, contain all these items and have recently also become
capable of producing disc galaxies with realistic structural parameters and
star formation histories (e.g. \citealp{a13, marinacci}) and have also been
used to study disc heating \citep{house, bird, martig}. However, the
uncertainties regarding the proper modelling of hydrodynamics and `sub-grid'
physics are still substantial (e.g.\ \citealp{aquila}), and simulations with
a resolution of $<100\pc$ as desired are still very costly, which prevents
large sets of such models.  Models of
idealised disc galaxies growing from the cooling of hot gas in an idealised
and isolated or cosmological dark matter halo \citep{roskar, aw13} are more
easily accessible for the problems considered here, but concerns regarding
hydrodynamics and sub-grid models and their effect on young stellar
populations in the simulations remain. For example, \citet{house} discuss how
birth velocity dispersion depends on resolution and the assumed star
formation prescription.

We therefore pursue a different, complementary approach, which allows us to study disc
heating with $N$-body simulations. We return to the approach pioneered by
\citet{selcar}, but used in only a few papers since then (notably the study by
\citealp{berrier} of the emergence of exponential surface density profiles from
arbitrary angular momentum distributions of infalling matter). In this approach
particles representing stars, and possibly gas and GMCs, are continuously
added on near-circular orbits. Spiral structure and bars are naturally
excited in the disc, and they both accelerate the stars and cause them to
migrate radially. We analyse our final models to
determine how they compare with observational diagnostics of disc heating in
the MW. Given the need to use arbitrary prescriptions for sub-grid physics in
full hydrodynamical simulations of disc growth, the models are no less
rigorous than full simulations, and, since they are computationally
cheaper, we can run large numbers of models and gain a reasonable
understanding of how a growing disc can evolve in isolation.

In this paper, we (i) present a set of novel numerical simulations 
of growing galaxies and explain how they were set up and run,
and (ii) give an overview of the most important lessons learnt
from these simulations. More detail will follow in papers on particular
aspects of Galactic growth, such as the age-velocity-dispersion relation for
nearby stars and its connection to the underlying heating laws, the distribution
function of the dark halo, the impact of spiral structure on star surveys or the 
details of radial migration processes. The structure of the paper is as
follows: Section 2 defines the simulations.
Section 3 justifies our modelling strategy by showing insensitivity to the
adopted initial conditions and adequate resolution. Section 4 demonstrates
the fundamental role played by GMCs, explores the more minor role played by
space-filling interstellar gas, and discusses heating and radial migration
within discs. A key result of this section, in which the disc mass and the
structure of the dark halo are in line with conventional wisdom, is that, while
thin discs very much like that of our Galaxy form in many simulations, if a
thick disc is required at the current epoch, it has to be added to the
initial conditions.  So in Section 5 we explore non-standard disc masses and
structures of the dark halo, and show that adopting a non-standard disc or halo,
while not solving the problem of thick-disc formation, spoils the good
agreement with observation found earlier as regards the bar and thin disc. 
In Section 6 we discuss the idealisations and potential missing ingredients 
of our models. Finally, in Section 7 we sum up and consider future work.

\section{Simulations}

We analyse a large set of controlled simulations of growing galactic discs
embedded in non-growing dark matter haloes. These simulations were all
carried out with the Tree Smoothed Particle Hydrodynamics (TreeSPH) code
GADGET-3, last described in \citet{gadget2}.  We apply different
gravitational softening lengths $\epsilon$ for baryonic and dark-matter
particles (listed as $\epsilon_{\rm DM}$ and $\epsilon_{\rm b}$ in Table
\ref{ictable}) and use an opening angle $\theta=0.5\,$degrees for the tree
code.  We use an adaptive time-stepping scheme in which timesteps are
fractions $\Delta_j\equiv\tau/2^j$ of a base timestep $\tau<10\,$Myr chosen
by the code: the $i$th particle is assigned timestep $\Delta_j$ when
\begin{equation}
\Delta_j<
\sqrt{2\eta \epsilon_i\over\left|{\bf{a}}_i\right|}<\Delta_{j-1},
\end{equation}
where ${\bf{a}}_i$ and $\epsilon_i$ are the particle's gravitational
acceleration and softening length, respectively, and $\eta=0.02$ is an
accuracy parameter. The hydrodynamical timestep is based on a Courant-like
condition $\Delta_{i,{\rm hyd}} \propto \kappa h_i/c_s$, where
$\kappa=0.15$ is the Courant parameter, $h_i$ is the smoothing length and
$c_s$ is the sound speed. We apply an isothermal equation of state $P=\rho
c_s^2$ and use $N_{\rm SPH}=48$ neighbours for the smoothing kernel. For
further code details we refer to \citet{gadget2}.

\subsection{Initial Conditions}

To generate initial conditions (ICs) for our numerical experiments we use the
publicly available GalIC code \citep{yurin}, which produces near-equilibrium
ICs of multi-component collisionless systems with given density distributions
using an iterative approach. 

\def\omitt{Starting from an initial assignment of particle
velocities, the difference of the time-averaged density response produced by
the particle orbits with respect to the initial density configuration is
characterised through a merit function, and a stationary solution of the
collisionless Boltzmann equation is found by minimising this merit function
directly by iteratively adjusting the initial velocities. Details of this
method are described in \citet{yurin}.}

The
initial systems consist of a dark halo with a mass in the range of
$M_{\rm{DM}}=(0.5-1)\times10^{12}\msun$ represented by
$N_{\rm{DM}}=(1-25)\times10^{6}$ particles and an embedded baryonic component.
The ICs feature an initial baryonic component in the form of a disc or a
bulge of mass $M_{\rm{b,i}}=(0.2-1.5)\times10^{10}\msun$ represented by
$N_{\rm{b,i}}=(2-15)\times10^{5}$ particles, so that the corresponding
baryonic particle masses are in the range $m_{\rm b}=(3.3-25)\times10^3\msun$.

The density of the dark halo is given by \citep{hernquist}
\begin{equation}
\rho_{\rm{DM}}(r)={{M_{\rm{DM}}}\over{2\pi}} {{a}\over{r\left(r+a\right)^3}}.
\end{equation}
The inner profile is adjusted so that it is similar to an NFW profile with
concentration $c_{\rm halo}=4-9$ and virial velocity $V_{200}\sim130-170\kms$.
The scale radii are in the range $a=24-52\kpc$. The halo initially has a
spherical density profile and radially isotropic kinematics, i.e. equal
velocity dispersion in the principal directions,
$\sigma_{r}=\sigma_{\phi}=\sigma_{\theta}$ and consequently  the
anisotropy parameter 
$\beta=1-\left(\sigma_{\theta}^2+\sigma_{\phi}^2\right) / \left(2
\sigma_{r}^2\right)=0$.

Our IC bulge components are set up with the same structure as those in \citet{yurin} and 
have a Hernquist density profile with scalelength $a_{\rm bulge}$ which is distorted 
to be mildly oblate with axis ratio $s=1.15$ as
\begin{equation}
\rho_{\rm bulge}(R,z)=s \rho_{\rm Hernquist}\left(\sqrt{R^2+s^2z^2}\right).
\end{equation}
We choose $a_{\rm bulge}\sim 500\pc$, which is smaller than the $800\pc$ found
by \citet{widrow} in their multi-component MW models with a Hernquist bulge,
as we would like to test the impact of compact bulge and as the MW bulge has 
likely grown over time due to secular processes. The bulge initially has no net rotation
and the velocity structure is axisymmetric.

Disc components are set up with a mass profile
\begin{equation}
{\rho_{\rm{disc,i}}(R,z)} = {{M_{\rm{b,i}}}\over{4\pi {z_{0,{\rm
disc}}}{h_{R,{\rm disc}}}^2}} {\sech^2 \left({z}\over{z_{0,{\rm
disc}}}\right)} {\exp\left(-{R}\over{h_{R,{\rm disc}}}\right)},
\end{equation}
with an exponential scalelength $h_{R,{\rm disc}}=1.5-2.5\kpc$ and a
radially constant isothermal vertical profile with scaleheights in the range
$z_{0, {\rm disc}}=0.1-1.2\kpc$. The vertical velocity dispersion $\sigma_z$ thus
declines with radius. For the ICs which start with a thin disc with
$z_{0, {\rm disc}}\sim0.1\kpc$, we assume ${\sigma_z^2}/{\sigma_R^2}=0.5$,
so that Toomre's $Q$ shows a minimum value of $Q_{\rm min}=1.15$ for our
standard halo.  For hotter IC discs we assume
${\sigma_z^2}/{\sigma_R^2}=1.0$.

For simulations with gas components we include a thin gas disc in the ICs.
For the YG and FG sets of ICs (see Table \ref{ictable}), the initial gas disc is created by turning a
randomly chosen 5 per cent of stellar particles in a Y or F IC into SPH
particles. We choose a low initial gas mass so that the disturbance of the 
IC is mild and unimportant compared to the rapid onset of disc instability in the 
growing galaxy. For the EG set of ICs we turn 1.66 per cent of star particles in the E
IC into gas particles, so that the initial gas mass is the same as in YG and FG.
For EG gas particles we choose only particles that lie close to the plane.

\begin{table*}
  \caption{An Overview over the different models and their parameters:
           {\it 1st Column}: IC Name;
           {\it 2nd Column}: Total IC mass $M_{\rm tot}$;
           {\it 3rd Column}: Total baryonic IC mass $M_{\rm b,i}$;
           {\it 4th Column}: Total gas IC mass $M_{\rm gas,i}$;
           {\it 5th Column}: Number of DM particles $N_{\rm DM}$ in IC;
           {\it 6th Column}: Number of baryonic particles $N_{\rm b,i}$ in IC;
           {\it 7th Column}: Concentration parameter for IC DM halo, $c_{\rm halo}$;
           {\it 8th Column}: IC DM halo scalelength $a_{\rm halo}$
           {\it 9th Column}:  IC radial disc scalelength $h_{R, {\rm disc}}$;
           {\it 10th Column}: IC vertical disc scaleheight $z_{0, {\rm disc}}$;
           {\it 11th Column}: IC bulge scalelength $a_{\rm bulge}$; 
           {\it 12th Column}: Gravitational softening length for DM particles, $\epsilon_{\rm DM}$; 
           {\it 13th Column}: Gravitational softening length for baryonic particles, $\epsilon_{\rm b}$; 
}
  \begin{tabular}{@{}ccccccccccccc@{}}\hline
    1st    & 2nd             &  3rd       & 4th         &5th           & 6th      & 7th          & 8th         & 9th             & 10th           & 11th   & 12th  & 13th\\
    {Name} &{$M_{\rm tot}$}    &{$M_{\rm b,i}$}&{$M_{\rm gas,i}$}&$N_{\rm DM}$   &$N_{\rm b,i}$&{$c_{\rm halo}$}&{$a_{\rm halo}$}&{$h_{R,{\rm disc}}$}&{$z_{0,{\rm disc}}$}&{$a_{\rm bulge}$}&{$\epsilon_{\rm DM}$}&{$\epsilon_{\rm b}$}  \\ 
           &{$[10^{12}\msun]$}      &{$[10^9\msun]$} & {$[10^8\msun]$} &             &           &              &{$[\kpc]$}    & {$[\kpc]$}       & {$[\kpc]$}   & {$[\kpc]$}   & {$[\pc]$}   & {$[\pc]$} \\ \hline
    Y      & $1$      &$5$&  --        &$5\,000\,000$ &$500\,000$ &    9        &  30.2       &  1.5           &   0.1            &   ---        & 134         & 30        \\\hline

    Z      & $1$      &$10$      &  --        &$5\,000\,000$ &$1\,000\,000$ &  9        &  30.2       &  1.5           &   0.5           &   ---        & 134         & 30      \\
    A      & $1$      &$10$      &  --        &$5\,000\,000$ &$1\,000\,000$ &  9        &  30.2       &  1.5           &   0.8           &   ---        & 134         & 30      \\
    E      & $1$   &$15$&  --        &$5\,000\,000$ &$1\,500\,000$ &  9        &  30.2       &  2.5           &   1.2           &   ---        & 134         & 30      \\
    EHR    & $1$   &$15$&  --        &$15\,000\,000$ &$4\,500\,000$ &  9        &  30.2       &  2.5           &   1.2           &   ---        & 90         & 20      \\\hline
    S      & $1$      &$2$&  --        &$5\,000\,000$ &$200\,000$ &    9        &  30.2       &  1.5           &   0.1            &   ---        & 134         & 30      \\\hline
    YLH    & $1$      &$5$&  --        &$1\,000\,000$ &$500\,000$ &    9        &  30.2       &  1.5           &   0.1            &   ---        & 200         & 30      \\
    YLR    & $1$      &$5$&  --        &$2\,000\,000$ &$200\,000$ &    9        &  30.2       &  1.5           &   0.1            &   ---        & 170         & 30      \\
    YHH    & $1$      &$5$&  --        &$25\,000\,000$ &$500\,000$ &    9        &  30.2       &  1.5           &   0.1           &   ---        & 80         & 30      \\ 
    YHR    & $1$      &$5$&  --        &$15\,000\,000$ &$1\,500\,000$ &    9        &  30.2       &  1.5           &   0.1        &   ---        & 90         & 20       \\\hline
    J      & $1$      &$5$&  --        &$5\,000\,000$ &$500\,000$ &    6.5      &  37.9       &  1.5           &   0.1            &   ---        & 134         & 30       \\
    JHR    & $1$      &$5$&  --        &$15\,000\,000$ &$1\,500\,000$ &    6.5      &  37.9       &  1.5           &   0.1            &   ---        & 90         & 20       \\
    F      & $1$      &$5$&  --        &$5\,000\,000$ &$500\,000$ &    4        &  51.7       &  1.5           &   0.1            &   ---        & 134         & 30       \\
    G      & $0.5$	&$5$&  --       &$5\,000\,000$ &$500\,000$ &    9        &  24.0       &  1.5           &   0.1            &   ---         & 134         & 30      \\\hline
    C      & $1$      &$5$&  --        &$5\,000\,000$ &$500\,000$ &    9        &  30.2       &  ---           &   ---            &   0.45        & 134         & 30       \\\hline

    YG     & $1$      &$5$&$2.5$ & $5\,000\,000$&$500\,000$&  9        &  30.2       &  1.5           &   0.1            &   ---        & 134         & 30     \\
    EG     & $1$   &$15$&$2.5$ & $5\,000\,000$ &$1\,500\,000$ &  9        &  30.2       &  2.5           &   1.2           &   ---        & 134         & 30  \\
    FG     & $1$      &$5$&$2.5$ & $5\,000\,000$ &$500\,000$ &    4        &  51.7       &  1.5           &   0.1            &   ---        & 134         & 30  \\\hline
\hline
  \end{tabular}
  \label{ictable}
\end{table*}

As far as in-plane kinematics of discs are concerned, we have modified the
GalIC code in the following way: \citet{yurin} assume that the in-plane
velocity dispersion are set by $\sigma_{\phi}^2=\left<v_{\phi}^2\right>
-\left<v_{\phi}\right>^2=\sigma_R^2$, which contradicts the epicycle
approximation, which for low $\sigma_R$ yields
$\sigma_{\phi}^2/\sigma_R^2\simeq0.5$.  This approximation is, however, only
expected to be valid for $\sigma_R\la10\kms$ (e.g. Fig.~4.16 in
\citealp{gd2}).  Indeed, initial conditions of discs set with
$\sigma_{\phi}=\sigma_R$ are unstable in the sense that
$\sigma_{\phi}^2/\sigma_R^2$ adjusts to a value which varies with $\sigma_R$,
with a global mean value $0.66$ for thin-disc ICs.  We set up the initial
conditions such that $\sigma_{\phi}^2/\sigma_R^2=0.66$ at all radii and apply
a phase mixing procedure to allow an adjustment to the correct radial
variation of ${\sigma_{\phi}^2}/{\sigma_R^2}$.  We take the output ICs of
GalIC and integrate each particle's equations of motion in the fixed
potential given by all particles at the output time until the radial
component of velocity, $v_r$, has changed from negative to positive values at
least eight times. We then select a random point in time during the
integrated time-span and adopt the corresponding phase space coordinates as
the new ICs. We find that we thus greatly suppress initial variations for
$\sigma_{\phi}$ and $\sigma_R$. For a MW-like, stable disc with $Q_{\rm
min}=2$ and $N_{\rm disc}=10^6$ we find that the dispersions at $R=5h_R$
change by $<10$ per cent over $5\Gyr$. The changes are still smaller at
smaller radii.

\subsubsection{Selected ICs}

Table \ref{ictable} provides an overview of all initial conditions. The first
row describes the most important initial condition, that labelled Y. It
comprises a halo of $10^{12}\msun$ distributed over $5\times10^6$ particles
with softening length $134\pc$. The dark matter halo of the MW is generally
expected to have a mass close to this value \citep{xue}.The halo's
scalelength implies a concentration $c=9$ close to that predicted for the MW
by cosmology \citep{zhao}.  Moreover, similar halo parameters were adopted
for the model of \citet{as15}, which fulfiled constraints on Snhd circular
velocity and dark matter mass within the solar radius. The stellar disc of
our standard initial condition Y is compact and thin and contains only a small
mass fraction (10 per cent for most models) of the final disc mass of the models.
It has scalelength $h_{R,{\rm disc}}=1.5\kpc$ and scaleheight $z_{0,{\rm disc}}=0.1\kpc$,
and it contains $5\times10^9\msun$ distributed over $5\times10^5$ particles
with softening length $30\pc$. The Y IC includes no gas and no bulge.

The purpose of the next three initial conditions listed in Table~\ref{ictable},
Z, A and E, is to test the impact of thicker initial discs. They
have discs that are twice and three times more massive than in the Y IC, with
correspondingly increased numbers of star particles. These more massive discs
are associated with greatly increased scaleheights, and in the case of E an
increased scalelength. IC EHR is a higher resolution version of IC E.

IC S below these in Table~\ref{ictable} differs from Y in
having  a mere $2\times10^9\msun$ in its disc.

In the next series of ICs in Table~\ref{ictable}, we vary the numbers of particles
in halo and/or disc. They are all based on ICs Y. YLR is a lower resolution version of Y, while
YHR is a higher resolution version. For YLH we reduce $N_{\rm DM}$, but keep
$N_{\rm b,i}$ the same and in YHH we increase $N_{\rm DM}$ for constant $N_{\rm b,i}$.

\begin{figure}
\centerline{\includegraphics[width=7cm]{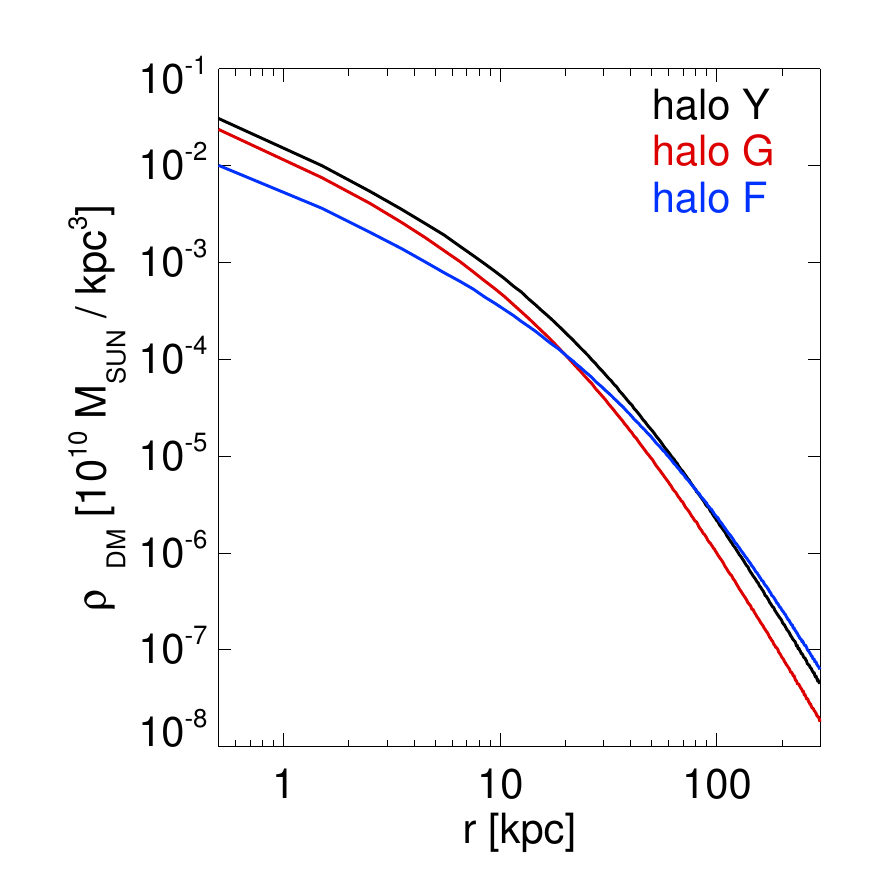}}
\caption{The radial density profiles of the standard dark halo (Y) and of two
non-standard dark haloes: that of F has a lower concentration, while that of
G has half the mass.}
\label{fig:aHalo}
\end{figure}

To create models with lower central dark matter densities,
the next group of ICs, J, F and G, differ from Y in the scalelengths $a_{\rm
halo}$ of their dark haloes. F has the standard halo mass but $a_{\rm halo}$
increased to $51.7\kpc$ from $30.2\kpc$, while G has only half the standard
mass in its halo but a small scalelength, $a_{\rm halo}=24\kpc$.  As
Fig.~\ref{fig:aHalo} demonstrates, increasing $a_{\rm halo}$ at fixed mass
lowers the density of dark matter in the central $\sim10 \kpc$, which is the
relevant region for our experiments, where the vast majority of stars live.
The halo of IC J is intermediate to those of ICs Y and F, while IC JHR is a
higher resolution version of J.

IC C differs from Y in having a bulge rather than a disc.

The final group of ICs in Table~\ref{ictable}, YG, EG, FG, are the only ICs to include
gas, $2.5\times10^8\msun$ of it. As their names suggest, YG results from
adding the gas to Y, EG results from adding gas to E, and FG is obtained by
adding gas to F.

\subsection{Feeding the disc}

\subsubsection{Adding stars}\label{sec:addstar}

To model the continuous growth of galactic discs via star formation, we add
new disc particles with masses $m_{\rm b}$ as determined by the baryonic
particle masses in the ICs to the existing disc every 5 Myr. For most models,
we assume a star formation rate (SFR) given by
\begin{equation}
  {\rm SFR} (t)= {\rm SFR}_0 \times \exp\left(-{{t}\over{t_{\rm SFR} }}\right),
\end{equation}
which we refer to as {\it type 1} SFR. The models use an exponential decay
timescale $t_{\rm SFR}=4-16\,{\rm Gyr}$, which is motivated by the findings
of \citet{ab09}. The specific choice of $t_{\rm SFR}$ in this range has a minor 
influence on the results presented here, which is why we focus on models with
$t_{\rm SFR}=8\,{\rm Gyr}$.  The normalisation ${\rm SFR}_0$ is always adjusted 
to yield a total inserted baryonic mass, including the disc mass from the ICs, 
of $M_{\rm f}\sim(3-8)\times10^{10}\msun$ after a time $t_{\rm f}=10\Gyr$
(\citealp{piffl} find $M_{\rm MW}=(5.6\pm1.6)\times10^{10}\msun$ for the MW).

For two models, we add an initial phase of increasing SFR through
\begin{equation}\label{eq:SFR}
  {\rm SFR} (t)= {\rm SFR}_0 \times \exp\left(-{{t}\over{t_{\rm SFR} }} - {{0.5 {\rm Gyr}}\over{t}}\right).
\end{equation}
We refer to this SFR law as {\it type 2} SFR. These models are evolved for $t_{\rm
f}=12$ Gyr and  the normalisation ${\rm SFR}_0$ is adjusted accordingly. When
modelling the chemical evolution of the disc, inclusion of the period in
which the SFR increases proves vital \citep{jason}. Since we do
not consider chemistry, the differences between models run with type 1 and
type 2 SFR prove to be modest. 

Note that due to the nature of our simulations, the final number of particles in the
disc $N_{\rm b,f}= M_{\rm f} / m_{\rm b} \gg N_{\rm b,i}$. For most simulations 
$N_{\rm b,f}\approx N_{\rm DM}$. 

The added disc mass is radially distributed as
$\Sigma(R)\propto\exp\left(-R/h_R(t)\right)$, with an exponential scale-length
that varies with time as 
\begin{equation}
h_R(t) = h_{R, {\rm i}} +  \left(h_{R, {\rm f}}-h_{R, {\rm i}}\right)
\left( t/t_{\rm f}\right)^\xi.
\end{equation}
Note that at $h_R(0)$ is not necessarily equal to the scale-length $h_{R,{\rm
disc}}$ of the IC. An increase in scalelength from $h_{R, {\rm i}}$ to $h_{R,
{\rm f}}$ at $t_{\rm f}$ for the pattern of accretion simulates inside out
growth of discs, which is suggested by observations of MW stellar
populations \citep{BensbyF, bovyrix} and external galaxies \citep{jing} and also by
cosmological models of disc galaxy formation \citep{a14}.  
\citet{bovyrix} find scalelengths in the range $h_R\sim1.5-5\kpc$
for their MW mono-abundance populations and we test different
radial growth histories within this range.
The particles start from $z=0$ and randomly chosen values of the azimuth $\phi$. 
The coordinate system is regularly updated to be centred on the centre of
mass of the system.

The particles are assigned near circular orbits. We determine the new
particles' rotational velocities as $v_{\phi}=\sqrt{{a_R(R)} R}$, where $a_R(R)$ is 
the azimuthal average of the radial gravitational acceleration,
$\partial\Phi/\partial R$.  To this value of $v_\phi$ we
add random velocity components in all three directions $\phi$, $R$ and $z$,
drawn from Gaussian distributions of dispersion $\sigma_0$. For most models,
$\sigma_0=6\kms$ is small and constant, in agreement with young local stars
\citep{ab09}. We have run one model with $\sigma_0=10\kms$ and found that it
does not change our results in any significant way. Motivated by observations
of turbulent discs at high redshift (e.g.\ \citealp{wisn}), for some models 
we assume that $\sigma_0(t)$ declines from a large initial value as
\begin{equation}
\sigma_0(t)=\left[ 6 + 30 \exp(-t / 1.5\Gyr) \right] \kms.
\end{equation}
Note that as we place all particles at $z=0$ the measured vertical dispersion
of an unheated component after insertion is smaller than $\sigma_0$, as the 
midplane is at the bottom of the vertical potential well and particles lose
kinetic energy moving away from it.

After a disc has developed a bar, there are no longer circular orbits
in the bar region. We can avoid this problem by introducing an inner cutoff
radius $R_{\rm cut}$, within which no particles are introduced. Our standard
approach is to set $R_{\rm cut}$ to the smaller of $5\kpc$ and the radius at which
the Fourier Amplitude of the stellar component
\begin{equation}
  A_2(R)\equiv{{1}\over{N(R)}} \sum\limits_{j=1}^{N(R)}e^{2\i\phi_j},
\label{bareq}
\end{equation}
 drops below $\e^{-1.5}$. Here $N(R)$ is the number of particles in a
thin cylindrical shell centred on $R$ and $\phi_j$ is the azimuth of particle $j$.
The majority of models are unaffected by the upper limit on $R_{\rm cut}$ as
few bars reach $5\kpc$. 

There are models for which our standard `adaptive' cutoff brings along
problems, which leads us to also use different approaches.
In the presence of an early strong bar, the adaptive cutoff very soon rises
to its limiting value, $5\kpc$. After a fraction of a gigayear $R_{\rm cut}$
decreases again. To avoid this early ring-like phase of growth, we have run
additional models with $R_{\rm cut}$ limited to $1\kpc$ during the first gigayear and
$3\kpc$ during the first $3\Gyr$. We refer to this as an `AdapLi cutoff'.

Unfortunately, adaptive cutoffs make radial star-formation histories
model-dependent. For some of our experiments, such as resolution tests
or comparisons of heating efficiencies, this effect makes the results harder to
interpret.  Therefore we use a `fixed' cutoff for certain subsets of our
models. If these models have type 1 SFR we fix the evolution of the cutoff
radius to
\begin{equation}
 R_{\rm cut}(t)=\left(0.67+{0.33t\over1\,{\rm Gyr}}\right) \kpc
\label{cutoff}
\end{equation}
at times $t>1\Gyr$. This rule mimics bar growth histories in test simulations with 
$R_{\rm cut}=0$ and no GMC particles. 

For model Y5fs4m2 (see Table \ref{modeltable2} and Section \ref{modelsrun} 
for details on the models and their names) which features 
a type 2 SFR, the cutoff is only introduced after $3\Gyr$ and then grows
in the same way starting at $1\kpc$. This reduces the impact of particles on
initially inappropriate orbits and also reduces the computational cost as
particles at greater radii on average have larger time-steps. Additionally,
it is not clear how the radial growth of galaxies happens in detail.
Cosmological simulations of forming disc galaxies by \citet{a14} find that
growing discs can display an inner region depleted of gas and star
formation. 

In addition to models with different cutoffs, we have also run a small 
number of test models, in which no cutoff was applied. As already mentioned, 
the choice of cutoff can influence the results of the models. 
When this is the case for the analysis of this paper, it is pointed out
in the text. In other cases, the conclusions are not affected by this choice.

\subsubsection{Adding and removing smooth gas}

The significance for
Galactic dynamics of the smoother component of the interstellar medium that
is the topic of this section is that, as a dissipative medium, it responds to
the non-axisymmetric part of the gravitational field in a different way to
stars. Our goal is to capture this essential difference with the simplest
possible model. We model gas with an isothermal equation of state $P=\rho
c_s^2$, where $c_s=10$ or $20\kms$ is the sound speed (\citealp{sormani2}
find that the effective sound speed in the ISM is $c_s\gtrsim10\kms$).  We grow the disc by
continuously adding star and gas particles. We achieve a roughly constant
global fraction $f_{\rm g}$ of the disc mass that comprises gas as follows: if the ratio of
gas mass to total mass in the disc exceeds $f_{\rm g}$, a fraction $f_{\rm g}$ of the
particles added to the disc in that timestep are gas particles; if the
gas-to-total mass ratio is less than $f_{\rm g}$, a fraction $2f_{\rm g}$ of the added
particles are gas particles. As our gas ICs have lower gas fractions than
the values of $f_g$ we use, gas simulations start with a period during which
$2f_{\rm g}$ of the added particles are gas particles.

To keep our models simple, we choose a constant gas fraction. Typical local star forming
disc galaxies have gas fractions $f_{\rm g}=0.05-0.25$ \citep{young}, whereas we test $f_g=0.1-0.3$.
Gas fractions have been shown to be higher at redshifts $z \sim2$ (e.g. \citealp{tacconi}).
However, we here only model the smooth interstellar gas. Our model for molecular clouds
is described in Section \ref{gmcsec} and produces, for typical parameters, a GMC fraction,
which declines from $f_{\rm GMC}>0.3$ at early times to $f_{\rm GMC}<0.05$ at late times.
We find that parameters $f_{\rm g}$ and $c_s$ in the ranges chosen for our models 
have a minor effect on our results.

High gas densities in the centres of model galaxies are computationally expensive, so
we limit the central gas density as follows. (i) We do not add gas
particles at $R<1\kpc$. Otherwise the same inner cutoff $R_{\rm cut}$ as for star
particles is applied.  (ii) We model star formation in the central galaxy by
identifying gas particles with hydrogen number densities $n>n_{\rm th}=10
\,{\rm cm}^{-3}$ and specific angular momentum $j_z<j_{\rm th}=100\kpc\kms$.
Such particles become the sites of star formation with a probability 
$p=1-\exp(-\nu \Delta t_i / t_{\rm dyn})$, where $\nu=0.1$ is an efficiency parameter, $t_{\rm
dyn}=1/\sqrt{4\pi G \rho}$ is the local dynamical timescale and $\Delta t_i$
is the timestep of the particle (e.g. \citealp{lia}). For each particle which is 
turned into a star, four gas particles which fulfil the star formation
criteria are removed from the simulation. This rule is designed to mimic the effect of a
central galactic outflow \citep{bland,M82}. In practice the above SF algorithm
comes into action only in the innermost $\sim2\kpc$, where gas densities are
high and material has low angular momentum. The region within which it is
active grows with increasing $f_{\rm g}$, decreasing $c_s$ and decreasing $h_R$.

As stated above, when the fraction of the disc mass that comprises gas falls
below $f_{\rm g}$, a fraction $2f_{\rm g}$ of added particles comprise gas. If the rate
of consumption of gas by SF in the central region is modest, adding a fraction
$2f_{\rm g}$ of gas particles causes the fraction of the disc mass that comprises
gas to rise until it reaches $f_{\rm g}$, when the fraction of added particles that
comprise gas drops to $f_{\rm g}$. Hence so long as the rate of consumption of gas
at the centre is modest, the gas fraction oscillates around $f_{\rm g}$. But for
a sufficiently large rate of central SF, even adding a fraction $2f_{\rm g}$ of gas
particles does not avert decline of the gas fraction.  Hence it can happen
that the gas fraction in the disc falls below $f_{\rm g}$, but it cannot rise
significantly above this value.

$M_{\rm f}$ (final baryonic mass) is defined to be the sum of the masses of
IC particles and all added particles, whether stars or gas. On account of the
ejection of gas by central SF, in models with gas the baryonic mass of the
final disc is always less than $M_{\rm f}$.

\tabcolsep=4.5pt
\begin{table*}
\vspace{-0cm}
  \caption{List of gas-free models; Models that contribute to a  figure have bold names.
           {\it 1st Column}: Model Name;
           {\it 2nd Column}: Initial Conditions;
           {\it 3rd Column}: GMCs Yes/No;
           {\it 4th Column}: Cutoff: No, Adaptive (no new particles in bar region), fixed (pre-defined evolving inner cutoff for new particles) or AdapLi (see text);
           {\it 5th Column}: Final time $t_{\rm f}$;
           {\it 6th Column}: Total inserted baryonic model mass $M_{\rm f}$ (including initial baryonic mass);
           {\it 7th Column}: Initial disc scalelength $h_{R, {\rm i}}$;
           {\it 8th Column}: Final disc scalelength $h_{R, {\rm f}}$;
           {\it 9th Column}: Scalelength growth parameter $\xi$;
           {\it 10th Column}: Type of SFR law;
           {\it 11th Column}: Exponential decay timescale $t_{\rm SFR}$ for the star formation rate;
           {\it 12th Column}: Initial velocity dispersion for inserted stellar particles, {$\sigma_0$};
           {\it 13th Column}: GMC star formation efficiency $\zeta$;
           {\it 14th Column}: GMC azimuthal density distribution parameter $\alpha$;
           {\it 15th Column}: GMC mass function lower and upper mass cutoffs {$M_{\rm low}$}, {$M_{\rm up}$};
           {\it 16th Column}: GMC mass function power law index;         
}
  \begin{tabular}{@{}cccccccccccccccc@{}}\hline
    1st   &  2nd  & 3rd   & 4th   &5th     & 6th         & 7th          & 8th            & 9th              & 10th    & 11th & 12th         & 13th        & 14th       & 15th       & 16th    \\
    {Name}&{ICs}  &{GMCs} &{Cutoff} &{$t_{\rm f}$}     &{$M_{\rm f}/\msun$}     &{$h_{R, {\rm i}}$} & {$h_{R, {\rm f}}$} & {$\xi$} & {SFR} & {$t_{\rm SFR}$}&{$\sigma_0$} & {$\zeta$}  & {$\alpha$} & {$M_{\rm low-up}$} & {$\gamma$}  \\ 
          &       &       &        &{$[\rm Gyr]$}&{$[10^{10}]$}    &{$\kpc$}        & {$\kpc$}        &	   & {type}& {$[\rm Gyr]$}&{$[\kms]$}  &            &            &	   {$[10^5\msun]$}     &        \\ \hline
   
    {\bf Y1}    &   Y    &  Yes  & Adap   & 10          &$5$&1.5             & 4.3             & 0.5     & 1     & 8.0          &  6         &0.08       & 1.0        & $1-100$       & $-1.6$  \\
    {\bf Y1f} 	    &   Y    &  Yes  & Fix    & 10          &$5$&1.5             & 4.3             & 0.5     & 1     & 8.0          &  6         &0.08       & 1.0        & $1-100$       & $-1.6$  \\ 
    Y1n             &   Y    &  Yes  & No     & 10          &$5$&1.5             & 4.3             & 0.5     & 1     & 8.0          &  6         &0.08       & 1.0        & $1-100$       & $-1.6$  \\   
    {\bf Y2}        &   Y    &  Yes  & Adap   & 10          &$5$&2.5             & 2.5             & 0.0     & 1     & 8.0          &  6         &0.08       & 1.0        & $1-100$       & $-1.6$  \\ 
    {\bf Y3}   	    &   Y    &  Yes  & Adap   & 10          &$5$&1.5             & 3.0             & 0.5     & 1     & 8.0          &  6         &0.08       & 1.0        & $1-100$       & $-1.6$ \\ 
    Y3f 	    &   Y    &  Yes  & Fix    & 10          &$5$&1.5             & 3.0             & 0.5     & 1     & 8.0          &  6         &0.08       & 1.0        & $1-100$       & $-1.6$ \\ 
    {\bf Y4} 	    &   Y    &  Yes  & Adap   & 10          &$5$&1.5             & 2.2             & 0.5     & 1     & 8.0          &  6         &0.08       & 1.0        & $1-100$       & $-1.6$ \\ 
    Y4f		    &   Y    &  Yes  & Fix    & 10          &$5$&1.5             & 2.2             & 0.5     & 1     & 8.0          &  6         &0.08       & 1.0        & $1-100$       & $-1.6$ \\ 
    {\bf Y5}  	    &   Y    &  Yes  & Adap   & 10          &$5$&1.5             & 3.5             & 0.5     & 1     & 8.0          &  6         &0.08       & 1.0        & $1-100$       & $-1.6$\\   
    Y6         &   Y    &  Yes  & Adap   & 10          &$5$&1.5             & 3.0             & 0.2     & 1     & 8.0          &  6         &0.08       & 1.0        & $1-100$       & $-1.6$  \\ 
    {\bf Y7}     	    &   Y    &  Yes  & Adap   & 10          &$5$&2.5             & 6.0             & 0.5     & 1     & 8.0          &  6         &0.08       & 1.0        & $1-100$  & $-1.6$  \\ \hline

    {\bf Y1Mb-}  &   Y    &  Yes  & Adap   & 10          &$3$&1.5             & 4.3             & 0.5     & 1     & 8.0          &  6         &0.08       & 1.0        & $1-100$       & $-1.6$  \\ 
    {\bf Y1Mb+}  &   Y    &  Yes  & Adap   & 10          &$7.5$&1.5           & 4.3             & 0.5     & 1     & 8.0          &  6         &0.08       & 1.0        & $1-100$       & $-1.6$  \\ 
    Y2Mb-   &   Y    &  Yes  & Adap   & 10          &$3$&2.5             & 2.5             & 0.0     & 1     & 8.0          &  6         &0.08       & 1.0        & $1-100$       & $-1.6$  \\ 
    Y2Mb+   &   Y    &  Yes  & Adap   & 10          &$7.5$&2.5           & 2.5             & 0.0     & 1     & 8.0          &  6         &0.08       & 1.0        & $1-100$  & $-1.6$  \\ \hline

    Y1s2            &   Y    &  Yes  & Adap   & 10          &$5$&1.5             & 4.3             & 0.5     & 1     & 16.0         &  6         &0.08       & 1.0        & $1-100$       & $-1.6$ \\
    Y1fs2           &   Y    &  Yes  & Fix    & 10          &$5$&1.5             & 4.3             & 0.5     & 1     & 16.0         &  6         &0.08       & 1.0        & $1-100$       & $-1.6$ \\
    Y1fs3    	    &   Y    &  Yes  & Fix   & 10          &$5$&1.5             & 4.3             & 0.5     & 1     & 4.0          &  6         &0.08       & 1.0        & $1-100$       & $-1.6$  \\ 
    Y4fs3    	    &   Y    &  Yes  & Fix    & 10          &$5$&1.5             & 2.2             & 0.5     & 1     & 4.0          &  6         &0.08       & 1.0        & $1-100$       & $-1.6$ \\ 
    Y5s4    	    &   Y    &  Yes  & Adap   & 12          &$5$&1.5             & 3.5             & 0.5     & 2     & 8.0          &  6         &0.08       & 1.0        & $1-100$       & $-1.6$ \\
    Y5fs4m2    	    &   Y    &  Yes  & Fix    & 12          &$5$&1.5             & 3.5             & 0.5     & 2     & 8.0          &  6         &0.08       & 1.0        & $1-30$	  & $-1.6$ \\
    Y5s5            &   Y    &  Yes  & Adap   & 12          &$5$&1.5             & 3.5             & 0.5     & 1     & 8.0          &  6         &0.08       & 1.0        & $1-100$ & $-1.6$  \\ \hline

    Y1f$\sigma$    &   Y    &  Yes  & Fix    & 10          &$5$&1.5             & 4.3             & 0.5     & 1     & 8.0          &  10        &0.08       & 1.0        & $1-100$  & $-1.6$  \\ \hline

    {\bf Y1f$\alpha$2} &   Y    &  Yes  & Fix    & 10          &$5$&1.5             & 4.3             & 0.5     & 1     & 8.0          &  6         &0.08       & 2.0        & $1-100$       & $-1.6$  \\
    {\bf Y1f$\alpha$0} &   Y    &  Yes  & Fix    & 10          &$5$&1.5             & 4.3             & 0.5     & 1     & 8.0          &  6         &0.08       & 0.0        & $1-100$       & $-1.6$  \\ 
    {\bf Y1f$\alpha$1} &Y    &  Yes  & Fix    & 10          &$5$&1.5             & 4.3             & 0.5     & 1     & 8.0          &  6         &0.08       & 1.5        & $1-100$   & $-1.6$  \\ \hline
   
    Y1$\zeta -$     &   Y    &  Yes  & Adap   & 10         &$5$&1.5             & 4.3             & 0.5     & 1     & 8.0          &  6         &0.04       & 1.0        & $1-100$       & $-1.6$  \\
    Y1l$\zeta -$    &   Y    &  Yes  & AdapLi & 10         &$5$&1.5             & 4.3             & 0.5     & 1     & 8.0          &  6         &0.04       & 1.0        & $1-100$       & $-1.6$  \\
    {\bf Y1f$\zeta -$} &   Y    &  Yes  & Fix    & 10          &$5$&1.5             & 4.3             & 0.5     & 1     & 8.0          &  6         &0.04       & 1.0        & $1-100$       & $-1.6$  \\
    Y1f$\zeta +$    &   Y    &  Yes  & Fix    & 10          &$5$&1.5             & 4.3             & 0.5     & 1     & 8.0          &  6         &0.16       & 1.0        & $1-100$       & $-1.6$  \\ 
    Y1f$\zeta +$m2  &   Y    &  Yes  & Fix    & 10          &$5$&1.5             & 4.3             & 0.5     & 1     & 8.0          &  6         &0.16       & 1.0        & $1-30$	  & $-1.6$ \\ 
    Y4$\zeta-$&   Y    &  Yes  & Adap   & 10          &$5$&1.5             & 2.2             & 0.5     & 1     & 8.0          &  6         &0.04       & 1.0        & $1-100$       & $-1.6$ \\ 
    {\bf Y4f$\zeta -$} &   Y    &  Yes  & Fix    & 10          &$5$&1.5             & 2.2             & 0.5     & 1     & 8.0          &  6         &0.04       & 1.0        & $1-100$  & $-1.6$ \\ \hline

    Y1m2            &   Y    &  Yes  & Adap   & 10          &$5$&1.5             & 4.3             & 0.5     & 1     & 8.0          &  6         &0.08       & 1.0        & $1-30$	  &  $-1.6$\\
    Y1fm2           &   Y    &  Yes  & Fix    & 10          &$5$&1.5             & 4.3             & 0.5     & 1     & 8.0          &  6         &0.08       & 1.0        & $1-30$	  & $-1.6$ \\  
    Y1nm2   	    &   Y    &  Yes  & No     & 10          &$5$&1.5             & 4.3             & 0.5     & 1     & 8.0          &  6         &0.08       & 1.0        & $1-30$	  &  $-1.6$\\
    {\bf Y1fm3}    &   Y    &  Yes  & Fix    & 10          &$5$&1.5             & 4.3             & 0.5     & 1     & 8.0          &  6         &0.08       & 1.0        & $10-10$      &$\delta$ fct.\\ 
    {\bf Y1fm4}    &   Y    &  Yes  & Fix    & 10          &$5$&1.5             & 4.3             & 0.5     & 1     & 8.0          &  6         &0.08       & 1.0        & $1-100$       & $-2.52$  \\ 
    {\bf Y1fm5}	    &   Y    &  Yes  & Fix    & 10          &$5$&1.5             & 4.3             & 0.5     & 1     & 8.0          &  6         &0.08       & 1.0        & $1-30$	  & $-1.78$ \\ 
    Y1fm6     	    &   Y    &  Yes  & Fix    & 10          &$5$&1.5             & 4.3             & 0.5     & 1     & 8.0          &  6         &0.08       & 1.0        & $1-100$  & $-2.2$ \\ \hline
 
    {\bf YN1 } 	    &   Y    &  No   & Adap   & 10          &$5$&1.5             & 4.3             & 0.5     & 1     & 8.0          &  6         & --         & --         & -- & --     \\
    {\bf YN1f}	    &   Y    &  No   & Fix    & 10          &$5$&1.5             & 4.3             & 0.5     & 1     & 8.0          &  6         & --         & --         & -- & --     \\ 
    YN2     	    &   Y    &  No   & Adap   & 10          &$5$&2.5             & 2.5             & 0.0     & 1     & 8.0          &  6         & --         & --         & -- & --     \\
    YN3     	    &   Y    &  No   & Adap   & 10          &$5$&1.5             & 3.0             & 0.5     & 1     & 8.0          &  6         & --         & --         & -- & --     \\
    YN3s2     	    &   Y    &  No   & Adap   & 10          &$5$&1.5             & 3.0             & 0.5     & 1     & 16.0         &  6         & --         & --         & -- & --     \\
    YN5s4     	    &   Y    &  No   & Adap   & 12          &$5$&1.5             & 3.5             & 0.5     & 2     & 8.0          &  6         & --         & --         & -- & --     \\
    YN7     	    &   Y    &  No   & Adap   & 10          &$5$&2.5             & 6.0             & 0.5     & 1     & 8.0          &  6         & --         & --         & -- & --     \\ \hline   
  \end{tabular}

  \label{modeltable}
\end{table*}

\begin{table*}
  \caption{Continuation of Table \ref{modeltable}.}
  \begin{tabular}{@{}cccccccccccccccc@{}}\hline
    1st    & 2nd   & 3rd   & 4th   &5th     & 6th         & 7th          & 8th            & 9th              & 10th    & 11th & 12th         & 13th        & 14th       & 15th       & 16th          \\
    {Name} &{ICs}  &{GMCs} &{Cutoff} &{$t_{\rm f}$}     &{$M_{\rm f}/\msun$}     &{$h_{R, {\rm i}}$} & {$h_{R, {\rm f}}$} & {$\xi$} & {SFR} & {$t_{\rm SFR}$}&{$\sigma_0$} & {$\zeta$}  & {$\alpha$} & {$M_{\rm low-up}$} & {$\gamma$}  \\ 
           &       &       &        &{$[\rm Gyr]$}&{$[10^{10}]$}
	   &{$\kpc$}        & {$\kpc$}        &         & {type}& {$[\rm
	   Gyr]$}&{$[\kms]$}  &            &            & {$[10^5\msun]$}     &        \\ \hline

    {\bf YLR1}& YLR& Yes  & Adap   & 10          &$5$&1.5             & 4.3             & 0.5     & 1     & 8.0          &  6         &0.08       & 1.0        & $1-100$       & $-1.6$  \\ 
    {\bf YHR1}& YHR& Yes  & Adap   & 10          &$5$&1.5             & 4.3             & 0.5     & 1     & 8.0          &  6         &0.08       & 1.0        & $1-100$       & $-1.6$  \\ 
    YHR2 &  YHR &  Yes  & Adap   & 10          &$5$&2.5             & 2.5             & 0.0     & 1     & 8.0          &  6         &0.08       & 1.0        & $1-100$  & $-1.6$  \\ \hline
    
    {\bf YLHN1f}&  YLH &  No   & Fix    & 10          &$5$&1.5             & 4.3             & 0.5     & 1     & 8.0          &  6         & --         & --         & -- & --     \\        
    {\bf YHHN1f}&  YHH &  No   & Fix    & 10          &$5$&1.5             & 4.3             & 0.5     & 1     & 8.0          &  6         & --         & --         & -- & --     \\
    YHRN1f&  YHR    &  No   & Fix    & 10          &$5$&1.5             & 4.3             & 0.5     & 1     & 8.0          &  6         & --         & --         & -- & --     \\\hline

    Z1              &   Z    &  Yes  & Adap   & 10          &$5$&1.5             & 4.3             & 0.5     & 1     & 8.0          &  6         &0.08       & 1.0        & $1-100$       & $-1.6$  \\ 
    Z1$\tau$        &   Z    &  Yes  & Adap   & 10          &$5$&1.5   & 4.3  & 0.5    & 1     & 8.0        &  $6 + 30\e^{-t_{1.5}}$&0.08       & 1.0        & $1-100$       & $-1.6$  \\ 
    {\bf Z2}        &   Z    &  Yes  & Adap   & 10          &$5$&2.5             & 2.5             & 0.0     & 1     & 8.0          &  6         &0.08       & 1.0        & $1-100$       & $-1.6$  \\
    ZN1             &   Z    &  No   & Adap   & 10          &$5$&1.5             & 4.3             & 0.5     & 1     & 8.0          &  6         & --         & --         &                  & --     \\
    ZN1$\tau$       &   Z    &  No   & Adap   & 10          &$5$&1.5   & 4.3  & 0.5    & 1     & 8.0        &  $6 + 30 \e^{-t_{1.5}}$& --         & --         &                  & --     \\\hline

    A1     &   A    &  Yes  & Adap   & 10          &$5$&1.5             & 4.3             & 0.5     & 1     & 8.0          &  6         &0.08       & 1.0        & $1-100$       & $-1.6$  \\ 
    A1$\tau$     &   A    &  Yes  & Adap   & 10          &$5$&1.5   & 4.3  & 0.5    & 1     & 8.0        &  $6 + 30\e^{-t_{1.5}}$&0.08       & 1.0        & $1-100$       & $-1.6$  \\ 
    {\bf A2}    &   A    &  Yes  & Adap   & 10          &$5$&2.5             & 2.5             & 0.0     & 1     & 8.0          &  6         &0.08       & 1.0        & $1-100$       & $-1.6$  \\
    {\bf A2$\tau$}	   &   A    & Yes   & Adap   & 10         &$5$ &2.5   & 2.5  & 0.0    & 1 &    8.0 &           $6 + 30\e^{-t_{1.5}}$ & 0.08 & 1.0 & $1-100$  & $-1.6$ \\
    AN1     &   A    &  No   & Adap   & 10          &$5$&1.5             & 4.3             & 0.5     & 1     & 8.0          &  6         & --         & --         &                  & --     \\
    AN1$\tau$     &   A    &  No   & Adap   & 10          &$5$&1.5   & 4.3  & 0.5     & 1     & 8.0        &  $6 + 30\e^{-t_{1.5}}$& --         & --         &                  & --     \\ \hline

    {\bf E1}     &   E    &  Yes  & Adap   & 10          &$5$&2.5             & 4.3             & 0.5     & 1     & 8.0          &  6         &0.08       & 1.0        & $1-100$       & $-1.6$  \\ 
    {\bf E1$\tau$}    &   E    &  Yes  & Adap   & 10          &$5$&2.5   & 4.3  & 0.5     & 1     & 8.0        &  $6 + 30\e^{-t_{1.5}}$&0.08       & 1.0        & $1-100$       & $-1.6$  \\ 
    {\bf E2}    &   E    &  Yes  & Adap   & 10          &$5$&2.5             & 2.5             & 0.0     & 1     & 8.0          &  6         &0.08       & 1.0        & $1-100$       & $-1.6$  \\
    EN1     &   E    &  No   & Adap   & 10          &$5$&2.5             & 4.3             & 0.5     & 1     & 8.0          &  6         & --         & --         &                  & --     \\
    {\bf EN1$\tau$}     &   E    &  No   & Adap   & 10          &$5$&2.5   & 4.3  & 0.5     & 1     & 8.0        &  $6 + 30\e^{-t_{1.5}}$& --         & --         &                  & --     \\
    EHR2 &  EHR & Yes  & Adap   & 10          &$5$&2.5             & 2.5             & 0.0     & 1     & 8.0          &  6         &0.08       & 1.0        & $1-100$       & $-1.6$  \\ \hline

    C1     &   C    &  Yes  & Adap   & 10          &$5$&1.5             & 4.3             & 0.5     & 1     & 8.0          &  6         &0.08       & 1.0        & $1-100$       & $-1.6$  \\ 
    {\bf C2}     &   C    &  Yes  & Adap   & 10          &$5$&2.5             & 2.5             & 0.0     & 1     & 8.0          &  6         &0.08       & 1.0        & $1-100$       & $-1.6$  \\ 
    CN1     &   C    &  No   & Adap   & 10          &$5$&1.5             & 4.3             & 0.5     & 1     & 8.0          &  6         & --         & --         &                  & --     \\\hline

    {\bf S1f}    &   S    &  Yes  & Fix    & 10          &$5$&1.5             & 4.3             & 0.5     & 1     & 8.0          &  6         &0.08       & 1.0        & $1-100$       & $-1.6$  \\
    S2     &   S    &  Yes  & Adap   & 10          &$5$&2.5             & 2.5             & 0.0     & 1     & 8.0          &  6         &0.08       & 1.0        & $1-100$       & $-1.6$  \\
    SN1     &   S    &  No   & Adap   & 10          &$5$&1.5             & 4.3             & 0.5     & 1     & 8.0          &  6         & --         & --         &                  & --  \\ \hline

    {\bf F1}     &   F    &  Yes  & Adap   & 10   &$5$&1.5             & 4.3             & 0.5     & 1     & 8.0          &  6         &0.08       & 1.0        & $1-100$       & $-1.6$  \\ 
    F1l     &   F    &  Yes  & AdapLi   & 10   &$5$&1.5             & 4.3             & 0.5     & 1     & 8.0          &  6         &0.08       & 1.0        & $1-100$       & $-1.6$  \\ 
    {\bf F2}    &   F    &  Yes  & Adap   & 10   &$5$&2.5             & 2.5             & 0.0     & 1     & 8.0          &  6         &0.08       & 1.0        & $1-100$       & $-1.6$  \\
    {\bf F2l}     &   F    &  Yes  & AdapLi   & 10   &$5$&2.5             & 2.5             & 0.0     & 1     & 8.0          &  6         &0.08       & 1.0        & $1-100$       & $-1.6$  \\
    {\bf F2l$\zeta -$} &   F    &  Yes  & AdapLi   & 10        &$5$&2.5             & 2.5             & 0.0     & 1     & 8.0          &  6         &0.04       & 1.0        & $1-100$       & $-1.6$  \\
    F3n     &   F    &  Yes  & No       & 10        &$5$&1.5             & 3.0             & 0.5     & 1     & 8.0          &  6         &0.08       & 1.0        & $1-100$       & $-1.6$  \\
    {\bf F3n$\zeta -$} &   F    &  Yes  & No       & 10        &$5$&1.5             & 3.0             & 0.5     & 1     & 8.0          &  6         &0.04       & 1.0        & $1-100$       & $-1.6$  \\
    FN1     &   F    &  No   & Adap   & 10   &$5$&1.5             & 4.3             & 0.5     & 1     & 8.0          &  6         & --         & --         &                  & --     \\
    FN1l    &   F    &  No   & AdapLi & 10   &$5$&1.5             & 4.3             & 0.5     & 1     & 8.0          &  6         & --         & --         &                  & --     \\
    FN2     &   F    &  No   & Adap   & 10   &$5$&2.5             & 2.5             & 0.0     & 1     & 8.0          &  6         & --         & --         &                  & --     \\
    FN2n    &   F    &  No   & No     & 10   &$5$&2.5             & 2.5             & 0.0     & 1     & 8.0          &  6         & --         & --         &                  & --     \\ \hline

    J1     &   J    &  Yes  & Adap   & 10          &$5$&1.5             & 4.3             & 0.5     & 1     & 8.0          &  6         &0.08       & 1.0        & $1-100$       & $-1.6$  \\ 
    J2     &   J    &  Yes  & Adap   & 10          &$5$&2.5             & 2.5             & 0.0     & 1     & 8.0          &  6         &0.08       & 1.0        & $1-100$       & $-1.6$  \\
    JN1    &   J    &  No   & Adap   & 10          &$5$&1.5             & 4.3             & 0.5     & 1     & 8.0          &  6         & --         & --         &                  & --     \\
    JHR2   &   JHR  &  Yes  & Adap   & 10          &$5$&2.5             & 2.5             & 0.0     & 1     & 8.0          &  6         &0.08       & 1.0        & $1-100$       & $-1.6$  \\\hline

    G1     &   G    &  Yes  & Adap   & 10          &$5$&1.5             & 4.3             & 0.5     & 1     & 8.0          &  6         &0.08       & 1.0        & $1-100$       & $-1.6$  \\ 
    G2     &   G    &  Yes  & Adap   & 10          &$5$&2.5             & 2.5             & 0.0     & 1     & 8.0          &  6         &0.08       & 1.0        & $1-100$       & $-1.6$  \\
    GN1  &   G    &  No   & Adap   & 10      &$5$&1.5             & 4.3             & 0.5     & 1     & 8.0          &  6         & --         & --         &                  & --     \\
    GN1l  &   G    &  No   & AdapLi   & 10      &$5$&1.5             & 4.3             & 0.5     & 1     & 8.0          &  6         & --         & --         &                  & --     \\ \hline
  \end{tabular}
  \label{modeltable2}
\end{table*}

\begin{table*}
\begin{center}
  \caption{Models that contain gas. All these models have $t_{\rm f}=10\Gyr$,
  $M_{\rm f}=5\times10^{10}\msun$, type 1 SFR, $t_{\rm SFR}=8\Gyr$ and
  $\sigma_0=6\kms$. The quantities given in cols 1--11 here are defined in
  the caption to Table~\ref{modeltable}. Cols.~12 and 13 give values for the
  gas-mass  fraction  $f_{\rm g}$ and the sound speed $c_s$. Models
  that contribute to a   figure have bold names.}
  \begin{tabular}{@{}ccccccccccccccccc@{}}\hline
    1st    & 2nd   & 3rd   & 4th   &5th     & 6th         & 7th          &    8th            & 9th              & 10th    & 11th & 12th & 13th  \\
    {Name} &{ICs}  &{GMCs} &{Cutoff}     &{$h_{R,{\rm i}}$} & {$h_{R, {\rm f}}$} & {$\xi$} & {$\zeta$}  & {$\alpha$} & {$M_{\rm low-up}$} &    {$\gamma$} & $f_{\rm g}$ & $c_s$ \\ 
           &       &       &        &{$\kpc$}	   & {$\kpc$}        &	   &            &            & {$[10^5\msun]$}     &       &	&	   $\!\kms$ \\ \hline

    {\bf YGN1}      &YG&No&Adap  &1.5             & 4.3             & 0.5     & --         & --         &                  & --  &$0.1$& $10$   \\
    YGN1c2      &YG&No&Adap&1.5             & 4.3             & 0.5     & --    & --         &                  & --    &$0.1$ & $20$   \\
    YGN1c2g2      &YG&No&Adap  &1.5             & 4.3             & 0.5     & --         & --         &                  & --   &$0.2$& $20$  \\
    YGN1c2g3      &YG&No&Adap  &1.5             & 4.3             & 0.5     & --         & --         &                  & --   &$0.3$& $20$  \\
    YGN3c2g2      &YG&No&Adap  &1.5             & 3.0             & 0.5     & --         & --         &                  & --   &$0.2$& $20$  \\ \hline
    {\bf YG1}     &YG&Yes&Adap &1.5             & 4.3             & 0.5     &0.08    & 1.0        & $1-100$       & $-1.6$ &$0.1$& $10$\\ 
    YG1c2      &YG&Yes&Adap &1.5             & 4.3             & 0.5     &0.08    & 1.0        & $1-100$       & $-1.6$ &$0.1$& $20$\\ 
    YG2g2      &YG&Yes&Adap &2.5             & 2.5             & 0.0     &0.08    & 1.0        & $1-100$       & $-1.6$ &$0.2$& $10$\\ 
    EG2      &EG&Yes&Adap &2.5             & 2.5             & 0.0     &0.08    & 1.0        & $1-100$       & $-1.6$ &$0.1$& $10$\\ 
    FG2f      &FG&Yes&Fixed &2.5             & 2.5             & 0.0     &0.08    & 1.0        & $1-100$       & $-1.6$ &$0.1$& $10$\\ \hline
  \end{tabular}
  \label{modeltable3}
\end{center}
\end{table*}

\subsection{Simulating GMCs}
\label{gmcsec}

Although GMCs are composed of gas, in our models, we treat them separately 
from the less dense gas that occupies much of the Galactic plane. 
To model the impact of a population of GMCs on the dynamics of the disc, we
introduce a population of very massive particles.  As stars form in GMCs the
total mass of GMCs present is determined by the SFR and the efficiency of
star formation $\zeta$, which gives the fraction of mass of a GMC that turns
into stars before supernovae completely destroy the GMC. For the MW,
\citet{murray} finds $\zeta=0.08$. GMCs are short-lived, with lifetime
$\tau_{\rm GMC}<100\Myr$:  \citet{murray} finds $\tau_{\rm GMC}\sim27\,{\rm
Myr}$ for massive GMCs in the MW and \citet{meidt} find a very similar value,
$\tau_{\rm GMC}\sim20-30\,{\rm Myr}$, for inter-arm clouds in M51.  However,
\citet{murray} also remarks that to account for the total molecular mass in
the MW one has to assume that molecular gas spends no more than half its time
in clouds, and the rest assembling and dispersing. 

We translate these facts into our models by adding a mass $\Delta m_{\rm GMC}
= \Delta m_{\rm stars} / \zeta$ in GMCs whenever we add a mass $\Delta m_{\rm
stars}$ in stars. A GMC lives for $\Delta t_{\rm GMC}= 50 \,{\rm Myr}$. The
mass of a GMC particle is a function of time:
\begin{equation}
m(t)=m_i\times
\begin{cases}
\zeta + t_{25}^2\left(1-{\zeta}\right)&t_{25}<1,\\
1&1<t_{25}<2,
\end{cases}
\end{equation}
where $t_{25}\equiv t/25\,\Myr$. Hence in the first half of its $50\Myr$
life, a cloud assembles from an initial mass $\zeta m_i$, and then its mass
is constant at $m_i$ until a burst of supernovae suddenly destroys it.
$\Delta m_{\rm GMC}$ is the total mass a coeval population of GMCs has 
reached after 25 Myr.

The typical sizes of massive GMCs in the MW are $l_{\rm GMC}\sim 10-100\pc$
(e.g.\ \citealp{murray}), close to the gravitational softening length
$\epsilon_{\rm disc}=30\pc$ of our disc particles in our standard resolution.
We therefore also model the GMC particles with $\epsilon_{\rm GMC}=30\pc$, 
which is advantageous because when particles with differing softening 
lengths interact, the code uses the longer length. When increasing (decreasing)
our resolution we decrease (increase) the softening lengths for
dark matter particles, but keep $\epsilon_{\rm GMC}$ the same. $\epsilon_{\rm disc}$
is only changed for increased resolution, when it is decreased.

GMC particles are introduced on near-circular orbits as described above for star
particles, assuming random velocity components with $\sigma_0=6\kms$. We have tested
setting $\sigma_0=0$ for GMCs and found no significant changes to the results. Their radial
distribution is the same as that of the stars added at the same time except
that we impose an outer cutoff at $R=R_{\rm cut}+6h_R$. The cutoff prevents
heavy GMC particles arising in the disc outskirts, where in real galaxies no dense
gas clouds are found.

In real galaxies, GMCs are preferentially found in higher-density disc
structures such as spiral arms and/or rings. To determine how the specific
spatial distribution of GMCs influences the heating process, we do not
distribute them uniformly in azimuth, as we do for star and gas particles.
Instead we make the density of added  GMCs be
\begin{equation}\label{eq:defalpha}
\rho_{\rm GMC}(R,\phi)\propto\rho_{\rm ys}^\alpha(R,\phi),
\end{equation}
where $\rho_{\rm ys}(R,\phi)$ is the density of young star particles with
ages between 200 and $400\Myr$, and $\alpha$ is a parameter. The youngest
stars are not included because they are introduced randomly in azimuth and
need time to fall in with structure in the disc, and older stars are excluded
as they are hotter and display weaker azimuthal structure. If $\alpha=0$, GMC
particles are added uniformly in azimuth, whereas if $\alpha>0$, any
structure displayed by young stars is reinforced by added GMCs. We determine
$\rho_{\rm ys}(R,\phi)$ by binning stars in 120 azimuthal and 36 radial bins,
with the radial extent of bins increasing with $R$.

Real GMCs have a spectrum of masses. Mass functions are usually
described as power-laws
\begin{equation}
{{\rd N}\over{\rd M}}\propto M^{\gamma}\hbox{ for }M<M_{\rm up}.
\end{equation}
Details of mass functions, such as $\gamma$ and the upper cutoff mass $M_{\rm
up}$ vary between galaxies and even within a galaxy (e.g.\ \citealp{roso,
paws}). In Local Group galaxies, $\gamma$ varies between $-1.4$ and $-2.9$
\citep{roso}, and thus between top-heavy and bottom-heavy distributions. The
most massive GMCs in the inner MW have $M\sim3\times10^{6}\msun$, which is
the number often used for $M_{\rm up}$ \citep{roso}, but
objects with masses close to $M\sim10^{7}\msun$ are also known \citep{murray,
garcia}, although they are usually considered to be cloud complexes. For M51,
the mass distributions have been measured to extend to  $M_{\rm
up}\sim10^{7}\msun$ \citep{paws}.

To account for these uncertainties, we ran models with a range of values for
$\gamma$ and $M_{\rm up}$. In most of our models we assume that all mass in
GMCs sits in distributions with a lower mass limit of $M_{\rm
low}=10^{5}\msun$. If the assumed mass function is not a delta function, we
sample it with eight different masses of GMC particles. For mass functions
with $M_{\rm up}=1\times10^7\msun$, the particle masses are
$\left\{0.25,0.5,1,2,4,6,8,10\right\}\times10^6\msun$ and for $M_{\rm
up}=3\times10^6\msun$ they are
$\left\{0.1,0.2,0.4,0.6,0.8,1,2,3\right\}\times10^6\msun$. We do not include
a radial variation of GMC mass function, such as is observed in the MW
\citep{roso}. 

\subsection{Models run}
\label{modelsrun}

Tables \ref{modeltable} and \ref{modeltable2} (which together form a single
long table) list 94 gas-free models we have run for at least $10\Gyr$ plus
model YHHN1f, which only ran until $9.3\Gyr$ due to a shortage of
resources.  We here describe these models and their naming convention.  Our
ten models that include an isothermal gas component are listed in Table
\ref{modeltable3}. Models that contribute to one of our figures are
marked with bold-face names.

These are all the models we currently have available. As the number is
limited by computational resources, and our rather large number of parameters
prevented us from a rigorous sampling of parameter space, we employed a two-fold strategy:
(i) We intended to understand how variations of parameters affect the results.
(ii) We intended to understand the phenomena displayed by all or a subset of the simulations.
For (ii) we thus explored some regions of parameter space more densely than others.

Model names all start with one or two capital letters identifying 
their initial condition as listed in Table \ref{ictable}.
Standard models have GMCs, but no gas. If there are no GMCs present, we add `N'
to the name. The presence of isothermal gas can be inferred from the added `G'
in the IC name.

These capital letters are followed by a number running from 1 to 7 describing
the radial growth history of the model, determined by parameters $h_{R, {\rm
i}}$, $h_{R, {\rm f}}$ and $\xi$. The most common growth histories are `1',
which stands for inside out growth starting from the initial condition $h_{R,
{\rm i}}=h_{R, {\rm disc}}$ and increasing as $t^{\xi=0.5}$ to $h_{R, {\rm
f}}=4.3\kpc$, and `2', which stands for a constant scalelength of $h_{R, {\rm
i}}=h_{R, {\rm f}}=2.5\kpc$. 

Each model name contains at least one capital letter for the IC and a number
for the growth history. For all other parameters we define standard values. Only
if a model deviates in one or more parameters from the standard, additional
digits are added to the model name as follows:

\begin{itemize}

\item Standard models use an adaptive inner cutoff $R_{\rm cut}(t)$. If we
use a fixed cutoff as described by equation \ref{cutoff}, we add `f' to the
name, if we use no cutoff we add `n' and for an AdapLi cutoff we add
`l'.

\item The vast majority of our models assume final baryonic mass $M_{\rm f}=5\times10^{10}\msun$.
We have two models each with $M_{\rm f}=3\times10^{10}\msun$ (labelled `Mb-')
and $M_{\rm f}=7.5\times10^{10}\msun$ (labelled `Mb+').

\item The overall star formation history (SFH) of a model is described by the 
SFR type, the final time $t_{\rm f}$ and the SF timescale $t_{\rm SFR}$. Our standard choice
is a type 1 SFR with $t_{\rm f}=10 \Gyr$ and $t_{\rm SFR}=8\Gyr$. We have explored
four additional SFHs, which are labelled by `s2',...,`s5'.

\item The standard input velocity dispersion $\sigma_0=6\kms$. We have one 
model with $\sigma_0=10\kms$ (labelled `$\sigma$') and several with 
$\sigma_0(t)=\left(6+30\e^{-t/1.5\Gyr} \right)\kms$ (labelled `$\tau$').

\item The GMC star formation efficiency $\zeta$ has a standard value of 0.08.
A lower $\zeta=0.04$ is labelled as `$\zeta-$', and a higher value of 
$\zeta=0.16$ is labelled as `$\zeta+$'.

\item $\alpha=1.0$ is the standard choice for the parameter controlling the azimuthal
density profile of GMCs. We have tested three differing values: $\alpha=0.0$ is labelled
`$\alpha0$', $\alpha=1.5$ is labelled `$\alpha1$' and $\alpha=2.0$ is labelled `$\alpha2$'.

\item The standard parameters controlling the GMC mass function are $M_{\rm low}=10^5\msun$,
$M_{\rm up}=10^7\msun$ and $\gamma=-1.6$. We have tested 5 additional mass functions, which
we label `m2',...,`m6'.

\item For the gas parameters our standards are $c_s=10\kms$ and $f_{\rm g}=0.1$.
      $c_s=20\kms$ is labelled as `c2', and higher values of $f_{\rm g}$ are labelled
      as `g2' and `g3'.
          
\end{itemize}

\begin{figure}
\centerline{\includegraphics[width=7cm]{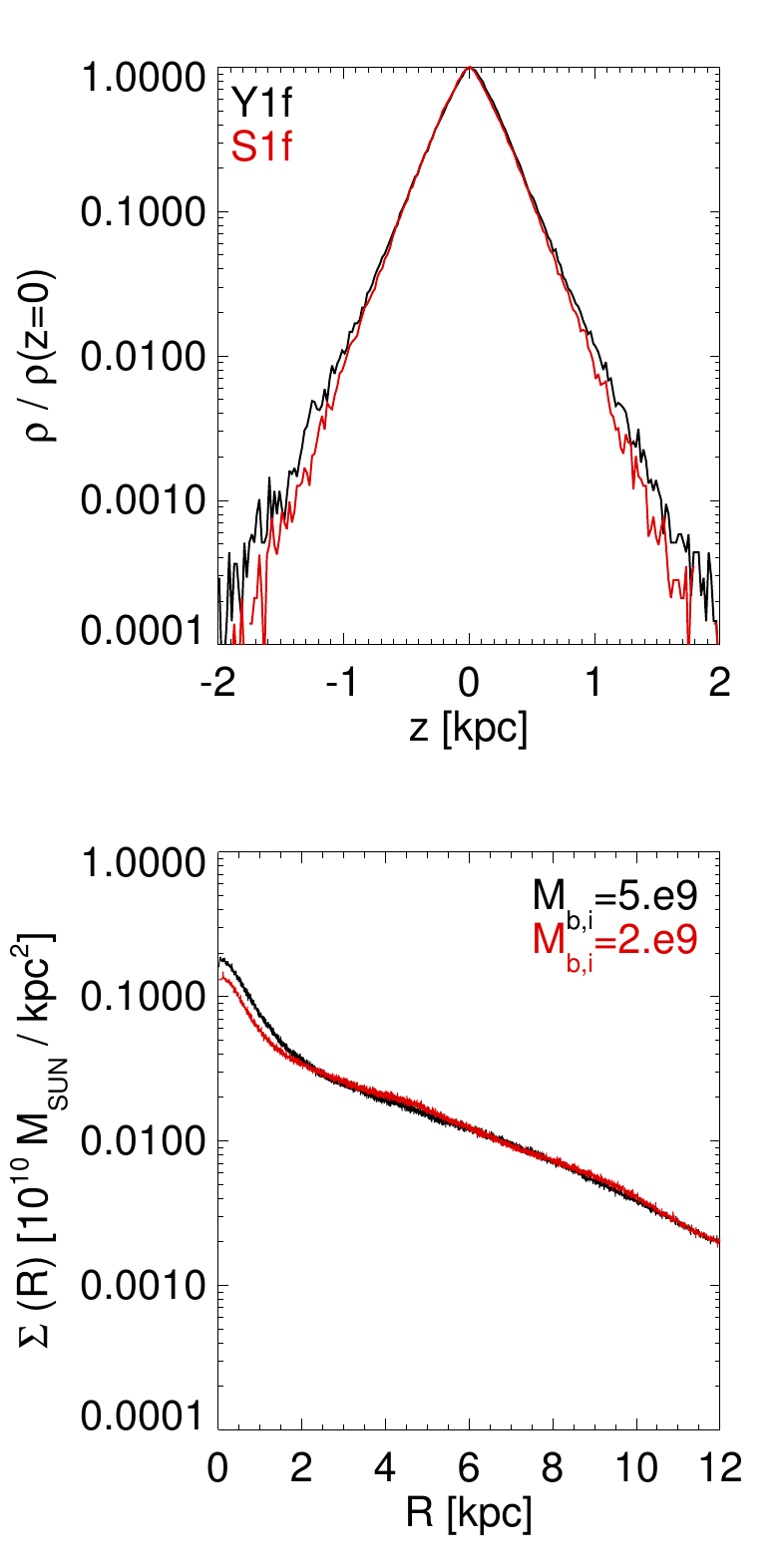}}
\caption{Comparison of models Y1f and S1f, which differ only in that the IC of
S1f has only 40 per cent of the stellar mass of Y1f's IC. Top: Vertical stellar density 
profile $\rho(z)$ at $R=8\kpc$. Bottom: Face-on stellar surface density profile $\Sigma(R)$.
}\label{fig:varyMIC}
\end{figure}

\section{Robustness of results}

Our simulations have value only if satisfactory answers can be given to two
questions: (i) how sensitive are the final models to details of the initial
conditions from which they are derived? and (ii) would the final models change
significantly if more particles were used to represent the galaxy? We now
address these questions. Unless mentioned otherwise, all results in this paper are presented
at final time $t=t_{\rm f}$ of the simulation and for results at $R=8\kpc$
we consider an annulus which is $1\kpc$ wide and centred on $R=8\kpc$.

\subsection{Impact of the initial baryonic mass}

Does the rather arbitrary mass $M_{\rm b,i}$ of our IC discs have a
significant impact on a model's evolution? Fig.~\ref{fig:varyMIC} addresses
this question by showing radial surface density profiles and vertical
density profiles at $R=8\kpc$ for two models, Y1f and S1f, that differ only
in that S1f's IC disc has only 40 percent of the mass of Y1f's IC disc.  At
fixed baryonic mass $M_{\rm f}$, a reduction in the mass $M_{\rm b,i}$ of the
IC has to be compensated by adding more mass later, using a longer
scalelength on average.  Consequently, decreasing $M_{\rm b,i}$ makes the
model less centrally concentrated, which weakens the central bar, if any. In
the lower panel of Fig.~\ref{fig:varyMIC}, the reduction in bar mass causes
the red curve of S1f to underlie the black one of Y1f at $R\la2\kpc$. We
shall see below that a weaker bar also accounts for the red curve underlying
the black one in the upper panel at $|z|\ga1\kpc$. However, from
Fig.~\ref{fig:varyMIC} it is clear that a 60 per cent reduction in the mass
of the initial disc changes the final model very little. Hence our models are
insensitive to $M_{\rm b,i}$. Comparison of models Y2/S2 and YN1/SN1
yields the same conclusion. As $M_{\rm b,i}$ is supposed to be small
compared to $M_{\rm f}$,  testing higher values of $M_{\rm b,i}$ in thin-disc
ICs would be
pointless. Models with $M_{\rm b,i}=0$ would be problematic because
an initial disc defines a disc plane, which is needed for insertion of
particles with (almost) parallel angular-momentum vectors.

A major finding of this study is that in our models a thick disc like that of
the MW must be present in the ICs, because it fails to form alongside the
thin disc (Section~\ref{sec:thkD}).  The insensitivity of the initial
conditions implies that major changes in the initial conditions are required
for the creation of a realistic thick disc.  Specifically, even a three-fold
increase in $M_{\rm b,i}$ and a tenfold increase in $z_{0,\rm disc}$ over the
values used in IC Y prove not quite sufficient to form a satisfactory
thick disc.

\subsection{A resolution study}
\begin{figure}
\centerline{\includegraphics[width=7cm]{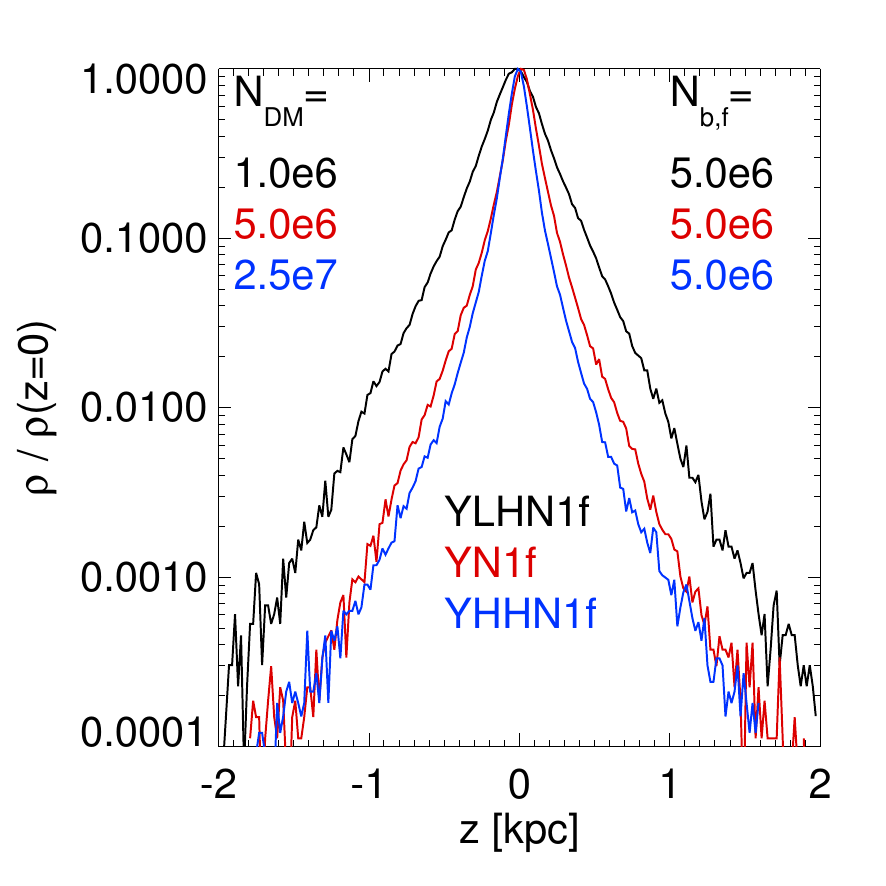}}
\caption{Vertical density profiles at $R=8\kpc$ of the stars of models
YLHN1f, YN1f and YHHN1f. These models differ only in the number $N$ of particles used to represent
that standard $M=10^{12}\msun$ dark halo: $N=1$, 5 and 25 million  for YLHN1f, YN1f
and YHHN1f, respectively. None of these models includes GMCs or gas. These results are presented
at $t=9.3\Gyr$, the last available output time for YHHN1f.
}\label{fig:resolveH1}
\end{figure}
\begin{figure}
\centerline{\includegraphics[width=7cm]{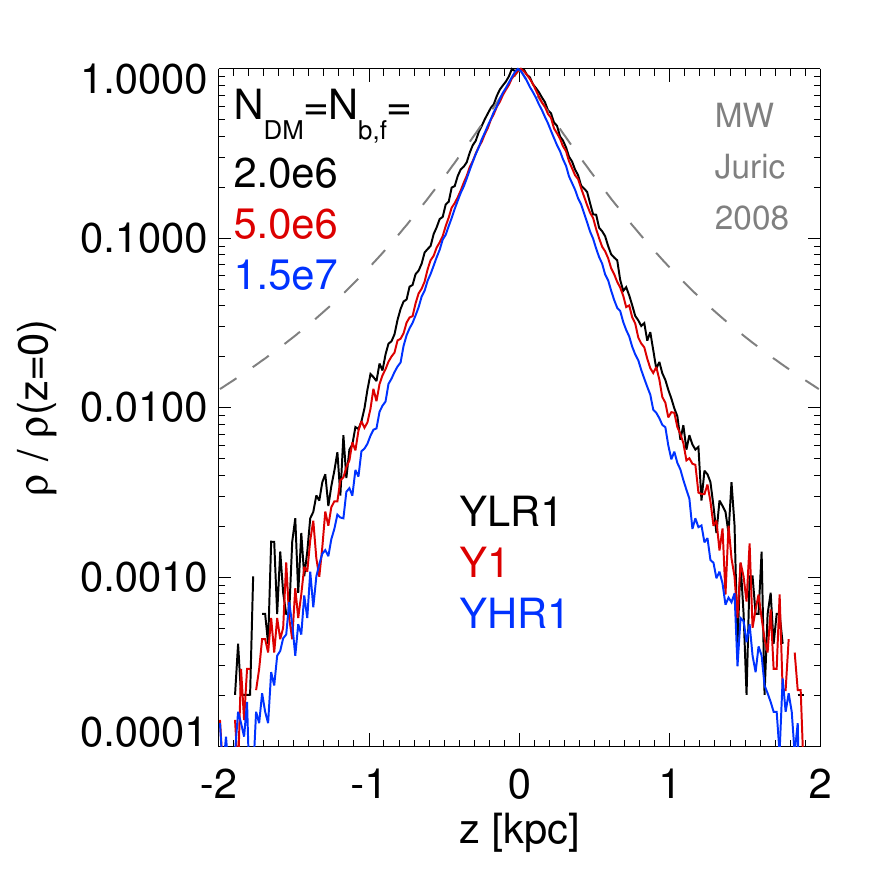}}
 \caption{As Fig.~\ref{fig:resolveH1} but showing models YLR1, Y1 and YHR1, which
contain GMCs and differ only in the numbers of particles used to represent
halo and stars: $(N_{\rm DM},N_{\rm b,f})=(2,2)$, $(5,5)$ and
$(15,15)\times10^6$ for YLR1, Y1 and YHR1, respectively.
The grey dashed lines show the vertical profile of the Galactic disc at the solar radius $R_0$ 
determined by \citet{juric}.
}\label{fig:resolveH2}
\end{figure}

Fig.~\ref{fig:resolveH1} shows the vertical density profiles at $R=8\kpc$ of
three models that contain neither GMCs nor gas and differ only in the number
$N$ of particles used to represent the dark halo.  Model YLHN1f has only one
million halo particles and as a consequence its disc is distinctly thicker
and hotter than the disc of model YN1f, which has five million particles in its
dark halo. This difference between the two discs is clearly attributable to
unphysical scattering of disc particles by halo particles. If we go to model
YHHN1f, which has 25 million particles in its halo, the resulting disc is only 
marginally thinner than the disc in YN1f.

Fig.~\ref{fig:resolveH2} shows an analogous plot for three models that all
contain GMCs but differ in the number of particles representing their dark
haloes and discs.  YLR1 has two million particles in the halo and two million
particles in the final disc, Y1 has five million particles each in halo
and final disc and YHR1 has 15 million particles each. 
Although the simulation with most particles (YHR1) clearly has the
thinnest disc, the three profiles do not differ much. In particular, the
differences are much smaller than that between the profiles plotted in
Fig.~\ref{fig:resolveH1} for models without GMCs. We conclude that when more
than a couple of million particles are used to represent the halo, spurious
two-body heating of the disc is swamped by physical heating by GMCs and is
consequently not a practical concern.

\section{Basic results}

In this section we summarise the most important lessons learnt from
simulations that employ the standard dark halo, being the halo predicted for
our Galaxy by concordance cosmology. In each subsection we explicitly mention
only a small subset of models that illustrate the relevant findings, but all
models have been analysed in the same way as the cited models, and the models
not mentioned are consistent with the stated findings. Further analysis
will be presented in future papers.

\subsection{Effectiveness of  GMC heating}
\label{seceff}
\begin{figure}
\centerline{\includegraphics[width=7cm]{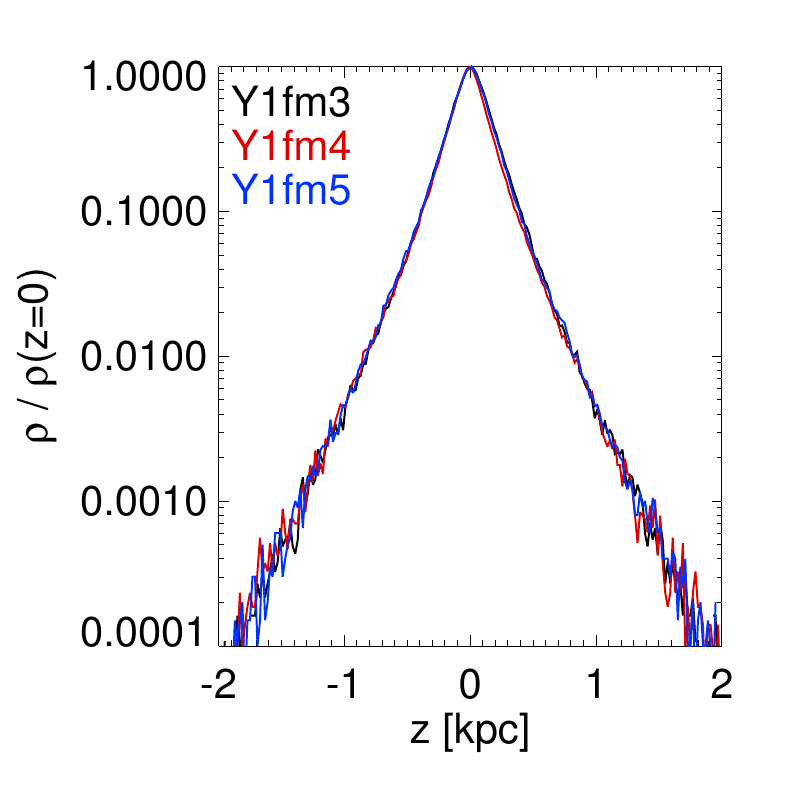}}
 \caption{The vertical stellar density profiles at $R=8\kpc$ of three models,
Y1fm3, Y1fm4 and Y1fm5 that differ only in their GMC mass functions.  Their
mass functions nevertheless yield the same effective mass
(eqn.~\ref{eq:effectM}) and as a consequence the models' vertical profiles
are indistinguishable.
}\label{fig:effectM}
\end{figure}

\begin{figure}
\centerline{\includegraphics[width=7cm]{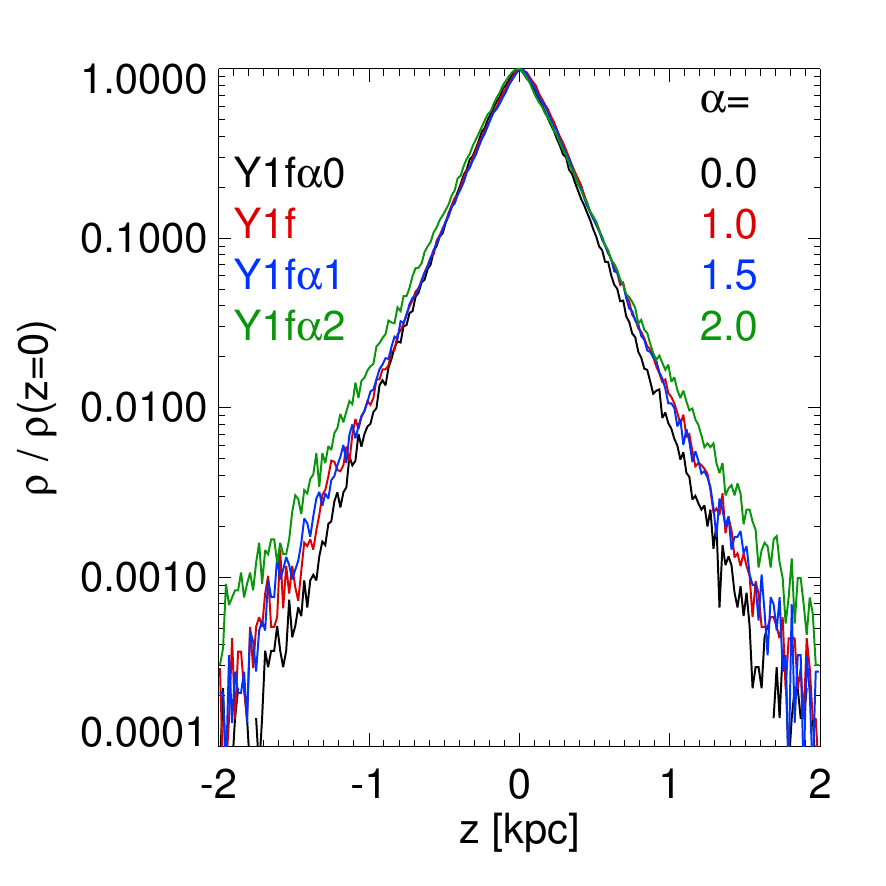}} 
 \caption{Vertical stellar density profiles at $R=8\kpc$ of four models with
different values of the parameter $\alpha$ that controls the extent to which
the density of GMCs is enhanced by spiral structure (eqn.~\ref{eq:defalpha}).
Concentrating GMCs (larger $\alpha$) into spiral arms enhances their heating
efficiency, but the effect is small.
}\label{fig:alpha}
\end{figure}
A remarkable feature of Fig.~\ref{fig:resolveH2} is that the vertical
profiles of all three models at $R=8\kpc$ are strikingly straight,
implying that GMC heating produces an accurately exponential vertical profile
out to at least $z=2\kpc$. Another remarkable fact is that the exponential
scale heights of these profiles ($h_z=210-235\pc$) are comparable to but
smaller than the measured scaleheight of the Galaxy's thin disc. 

The grey dashed lines in Fig.~\ref{fig:resolveH2} show
the vertical profile of the Galactic disc at the solar radius $R_0$ in the
analysis of \citet{juric} of the SDSS photometry. This profile is shown
in several figures throughout the paper and represents a bias-corrected
model-fit to the data of the form 
\begin{equation}
\rho(z, R=8\kpc)=\rho_0[\exp(-|z|/h_{\rm thin})+f\exp(-|z|/h_{\rm thick})].
\end{equation}
\citet{juric} find $h_{\rm thin}=300\pc$ and $h_{\rm thick}=900\pc$ with 
20 per cent uncertainty each and $f=0.12$ with 10 per cent uncertainty.

We see that GMCs produce a
disc that has a suitable scaleheight near the plane ($|z|\la0.5\kpc$) but
completely fails to fit the data at greater heights because it lacks a
thick-disc component.

\begin{figure*}
\centerline{\includegraphics[width=14cm]{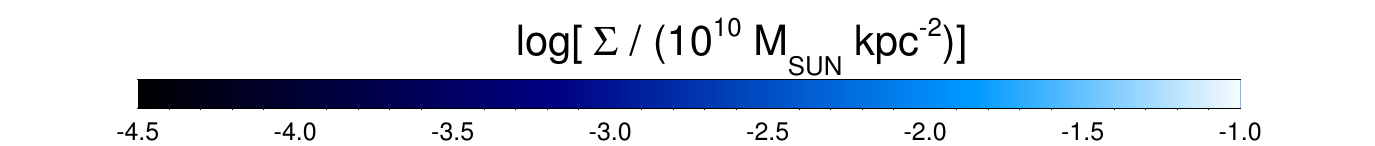}}
\centerline{\includegraphics[width=16cm]{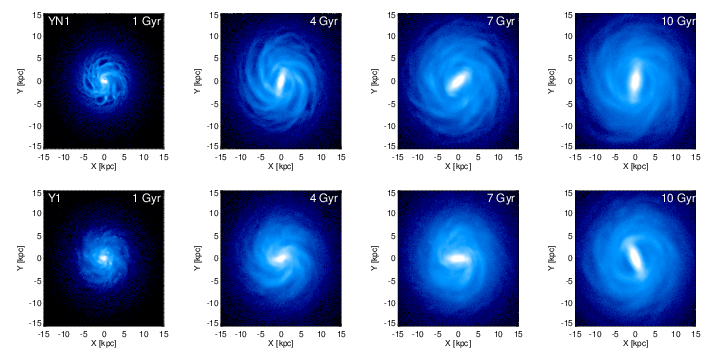}}
 \caption{Face-on stellar surface density $\Sigma(x,y)$ pictures of models
YN1 and Y1 at $t=1,4,7,10\Gyr$. }\label{fig:killBar}
\end{figure*}

The heating efficiency of GMCs depends on the contribution of the GMCs to
the disc's surface density, which is controlled by the SFR and the efficiency
parameter $\zeta$ -- smaller values of $\zeta$ imply a higher surface density
from GMCs at a given SFR. The heating efficiency also depends on the mass
spectrum of the clouds. \citet{jenkins} gave an analytic argument 
that the heating efficiency of clouds depends on the mass function $N(m)$
through the effective mass
\begin{equation}\label{eq:effectM}
\mathcal{M}\equiv{\int\rd m\, m^2N(m)\over\int\rd m\,mN(m)}.
\end{equation}
In Fig.~\ref{fig:effectM} we plot the vertical profiles of three models Y1fm3, Y1fm4 and Y1fm5 at $R=8\kpc$,
that all have effective mass $\mathcal{M}=10^6\msun$. We see that
the density profiles of these models coincide perfectly, as \citet{jenkins}
predicted.

In Fig.~\ref{fig:alpha} we explore the importance of concentrating GMCs in
spiral arms by increasing the parameter $\alpha$ in equation
(\ref{eq:defalpha}) from zero (no concentration) to 2 ($\rho_{\rm
GMC}\propto\rho_{\rm ys}^2$). As expected, concentrating GMCs into spirals
(as seen in observations, e.g. \citealp{schinnerer})
enhances the heating efficiency of GMCs by enabling them to act
cooperatively. The effect is, however, small. We will henceforth adopt
$\alpha=1$.

GMC heating is very effective at early times, because then (i) the SFR is
high so the surface density of GMCs is large, and (ii) an individual GMC
represents a larger fraction of the disc mass than it does today, when the
disc is ten times more massive than it was at $z=2$. An important effect of
prompt heating by GMCs is to delay, and in extreme cases cancel, the
formation of a bar. Fig.~\ref{fig:killBar} illustrates this point by showing
surface-density maps at four times for models YN1 (top), which has no GMCs, and
its GMC possessing partner Y1. YN1 has a prominent bar already at $4\Gyr$
while in Y1 the bar appears first at $7\Gyr$. Moreover, at $10\Gyr$ the bar
is longer in YN1 than in Y1.

To assess the strength and length of bars in our simulations we use equation
\eqref{bareq} to determine the profile of the $m=2$ Fourier amplitude $A_2$. We define
a bar as a central structure with $\ln(A_2)>-1.5$ and its length $L_{\rm bar}$ as the
radius where $A_2$ drops below this value. To determine the formation time of a bar
we plot $L_{\rm bar}$ as a function of time $t$ and look for the earliest time after 
which $L_{\rm bar}$ is continuously larger than 1 kpc. Note that bars can dissolve and reform.

The MW bar has a distinct `X' shape \citep{wegg}, as have many bars in edge-on galaxies.
This is the result of the buckling instability \citep{combes}, which occurs when the 
bar has acquired a critical strength. As far as the bars in our simulations are concerned,
we find that in the absence of GMCs, or in models with low GMC effective mass 
$\mathcal{M}$, all bars buckle. If we consider different Y1 models irrespective of the 
cutoff prescription and define an efficiency parameter $\chi=\mathcal{M}/(10^6\zeta\msun)$,
all models with $\chi<14$ buckle. For $14<\chi<40$ we find both buckled and unbuckled
bars, as the evolution of bar length and strength is to some degree stochastic
\citep{sellwood09} and the detailed evolution history of a model bar decides if
a bar buckles or not. All models with $\chi>40$ have weak bars, which do not buckle.
The bars in models YN1 and Y1 shown in Fig.~\ref{fig:killBar} have both buckled.

\subsection{AVR}

\begin{figure*}
\centerline{\includegraphics[width=7cm]{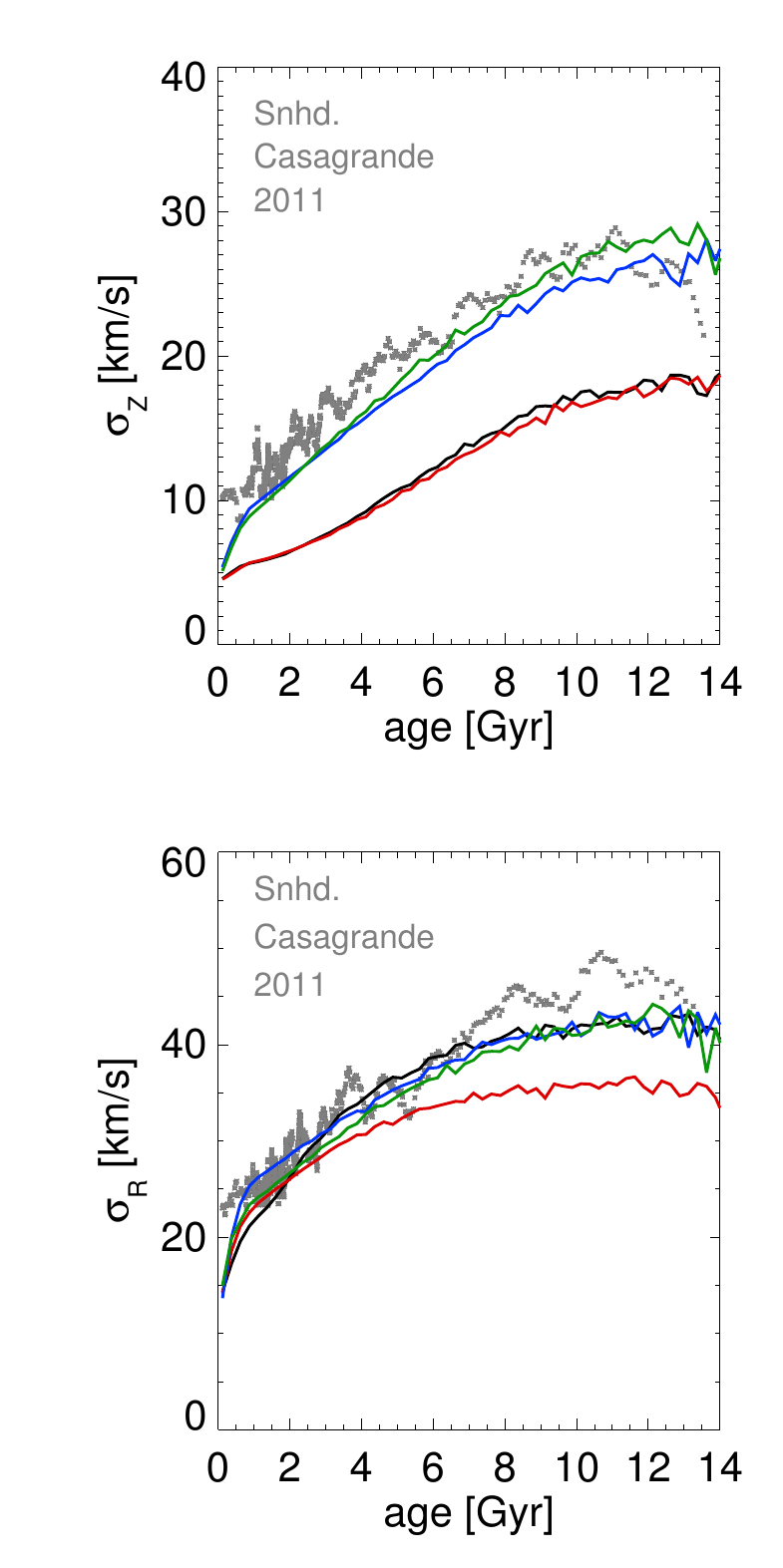}
\includegraphics[width=7cm]{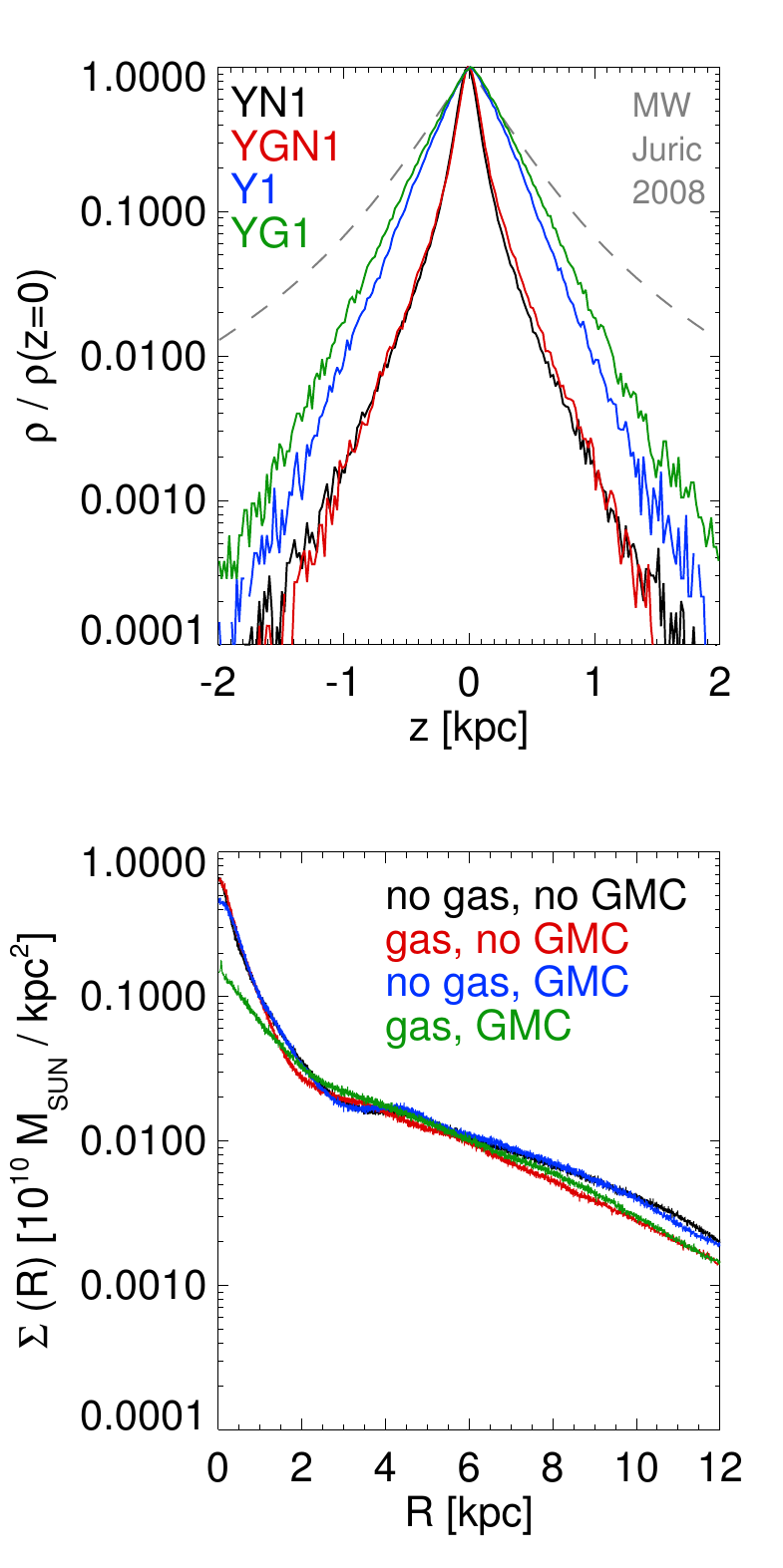}
}
 \caption{A comparison of models YN1 (no gas, no GMCs), YGN1 (gas, no GMCs),
Y1 (GMCs, no gas) and YG1 (gas, GMCs). Top right: vertical stellar density
profile $\rho(z)$ at $R=8\kpc$. Bottom right: radial stellar surface density
profiles $\Sigma(R)$. Left hand panels: Age velocity-dispersion ($\sigma_z$
upper, $\sigma_R$ lower panel) relations of stars at $R=8\kpc$.  The
grey dots show data for the Snhd from Casagrande et al.~(2011) and the model
curves have been adjusted to simulate the impact on measured dispersions of
the uncertainties and biases in the ages determined by Casagrande et
al. 
}\label{fig:gasGMC}
\end{figure*}

In the left hand panels of Fig.~\ref{fig:gasGMC} we compare $\sigma_z$ (upper panel)
and $\sigma_R$ (lower panel) at $R=8\kpc$ as a function of stellar age $\tau$ at the endpoints 
of four models that differ only in whether or not they have GMCs (Y1 and YG1 do) and whether
or not they have gas (YG1 and YGN1 do). We overplot in grey $\sigma_i(\tau)$ as 
measured for the Snhd by \citet{casagrande}. For these points, which are
shown in several plots throughout the paper, only stars with 
relatively low $1\sigma$ age errors $\sigma_{\tau}$ are considered
($\sigma_{\tau}<1\Gyr$ or $\sigma_{\tau} /\tau< 0.25$). Moreover, stars with
$\feh<-0.8$ or $V<-150\kms$ are excluded to avoid contamination with halo stars.

The major uncertainties in the Snhd AVR for very old stars are connected to the
uncertainties in excluding halo stars. Fig. 17 in \citet{casagrande} illustrates
this and shows that these uncertainties are $\sim 10$ per cent. For young stars 
uncertainties in the AVR derive from the age errors and the intrinsic shape of 
the AVR. Generally, the age uncertainty in the Casagrande et al. data is 
substantial and can signficantly modify $\sigma(\tau)$ by contamination 
from neighbouring age bins. Moreover, the GCS selection function excludes very
blue stars and thus removes stars from the sample that could be safely identified
as very young ($\tau \lesssim 1 \Gyr$) objects. It also favours greatly stars with
ages $\tau \sim 2 \Gyr$, so that the lowest age bins are dominated by stars with 
age-underestimates from that peak in the age-distribution. One can use the bluest 
\emph{Hipparcos} stars for which \citet{ab09} find $\sigma_z=5.5\kms$ and 
$\sigma_R=8\kms$ to estimate the dispersions of the youngest stars,
which are clearly smaller than what we find from the youngest GCS stars.
However, these stars may be kinematically biased as they belong to a low number
of moving groups of young stars. Also, these blue stars have associated ages which 
are smaller than the typical age errors of observed stars and smaller than the bin
sizes used for the simulations, so that neither for the simulations nor for 
observations we would expect to recover the velocity dispersion of this population. 

To attenuate some of these uncertainties, the model curves plotted in Fig.~\ref{fig:gasGMC}
have been adjusted to allow for the impact of the Casagrande et al.\ age uncertainties 
and biases (Aumer et al.\ in preparation). Selection function effects for the bluest 
stars are, however, not taken into account. Thus our simulations overestimate
the number of intrinsically very young stars and hence underestimate the relative 
contamination with age underestimates. This explains why our model dispersions are lower
than observed dispersions at $\tau \lesssim 1 \Gyr$. As the stars examined by Casagrande 
et al.\ were all observed close to the Galactic  midplane, for our models we only 
consider stars at low altitudes $|z|<100\pc$. Furthermore, the simulations only cover 
10 Gyr and we assign ages of 10-11 Gyr to the IC particles. So at ages $\tau>11\Gyr$, 
the simulation curves only show upscattered younger stars, whereas the Snhd data also 
contains stars which are actually older than 11 Gyr.

Considering these complications, we find that the models with
GMCs provide moderate fits to the data for $\sigma_z(\tau)$, while the models
without GMCs have $\sigma_z$ too small.

The absence of GMCs from YN1 does not stop this model providing a moderate fit
to the data for $\sigma_R(\tau)$, but adding gas to this model to make YGN1
does push $\sigma_R(\tau)$ well below the data. We explain this phenomenon in the
next section. The fact that GMCs are not needed to explain observed $\sigma_R(\tau)$,
but are required to explain $\sigma_z(\tau)$ confirms \citet{carlberg}, who argued
that heating by spirals and bars can explain the in-plane dispersions, but GMCs 
are needed to explain the vertical heating.

\subsection{Impact of gas}

The right hand panels of Fig.~\ref{fig:gasGMC} compare vertical stellar profiles at $R=8\kpc$ (upper panel) 
and radial stellar profiles (lower panel) of the four models YN1, YGN1, Y1 and YG1, discussed in
the last section.  The black and red vertical profiles for the GMC-free
models YN1 and YGN1 are essentially identical (as are their curves for
$\sigma_z(\tau)$), so on its own, gas is not an effective heating agent.
Since the green curves for YG1 lie above the blue curves for Y1 in the upper
panels of Fig. \ref{fig:gasGMC}, we conclude
that gas \emph{does} enhance the heating efficiency of GMCs. This is
presumably because a GMC particle attracts a wake of gas particles, and this wake increases the
GMC's effective mass.  In Fig.~\ref{fig:gasGMC} the vertical profiles of the
models with GMCs are almost perfectly exponential with scaleheights $h_z=225$
(no gas) and $253\pc$ (with gas).

\begin{figure*}
\centerline{\includegraphics[width=14cm, angle=90]{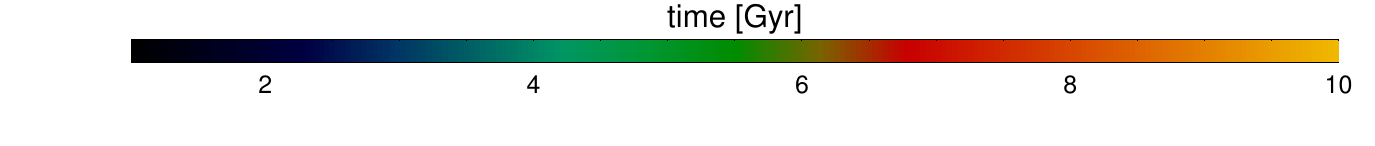}
\includegraphics[width=7cm]{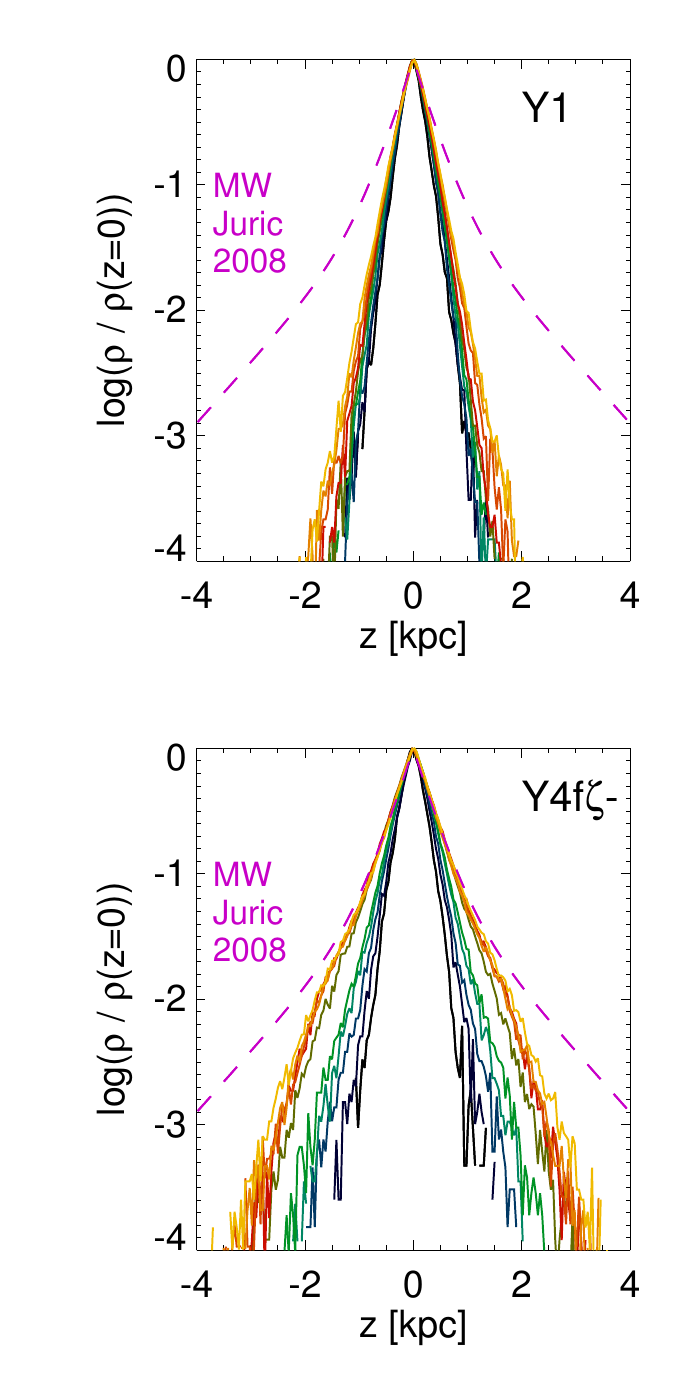}
\includegraphics[width=7cm]{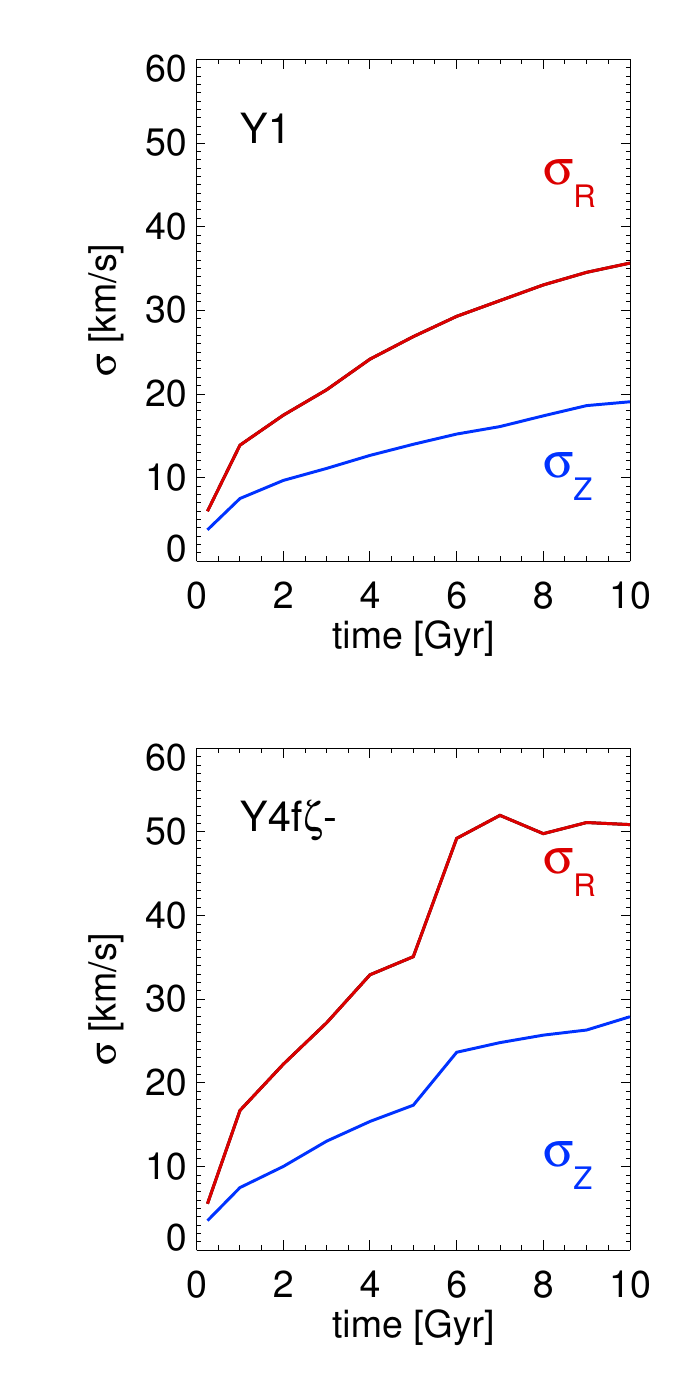}}
 \caption{Right hand panels: velocity dispersion $\sigma_z$ at $R=8\kpc$ as a
function of time in models Y1 and Y4f$\zeta-$. Left hand panels: Time
evolution of vertical stellar density profiles at
$R=8\kpc$. The colour bar at extreme left shows the time encoding.
}\label{fig:rhozt}
\end{figure*}

The radial profiles shown in the lower right panel of Fig.~\ref{fig:gasGMC}
comprise a steep section for $R<2\kpc$ associated with the bar, flattening
into an extended outer section for the disc. Model YG1, with both gas and
GMCs, has a distinctly less prominent bar section than the other three
models. We have already seen that GMCs tend to weaken bars. The profile for
YG1 signifies that gas also weakens bars. Gas can weaken bars in three ways.
One is simply by enhancing the effectiveness of GMC heating.  Another is by
blowing mass out of the central galaxy, thus weakening the disc's
self-gravity. The third way gas can weaken a bar is by
carrying angular momentum over the bar's corotation resonance and then
surrendering it to the bar: when a bar acquires angular momentum, it becomes
faster, shorter and weaker -- and conversely when it loses angular momentum
to the dark halo. The first process will not be active in model YGN1 that
has gas but no GMCs, and the radial profile in Fig.~\ref{fig:gasGMC}
indicates that the bar in this system is significantly shorter than that in
YN1, which has neither gas nor GMCs. Thus on its own gas does not prevent bar
formation, but it does limit bar growth. Acting in concert with GMCs, gas can
strongly delay bar formation.

Note that the radial profiles show only stellar mass. As gas models have a fraction
of their disc mass in gas, their overall stellar surface density is lower.
Still, at large $R$ the radial profiles of the two models without gas fall off
less steeply than the profiles of the models with gas. That is, adding gas reduces the
scalelength of the final disc. Presumably, this  is partly a side effect of
reducing the strength of the bar, and partly caused by viscous inflow of gas
through the disc.

\begin{figure*}
\centerline{\includegraphics[width=12cm]{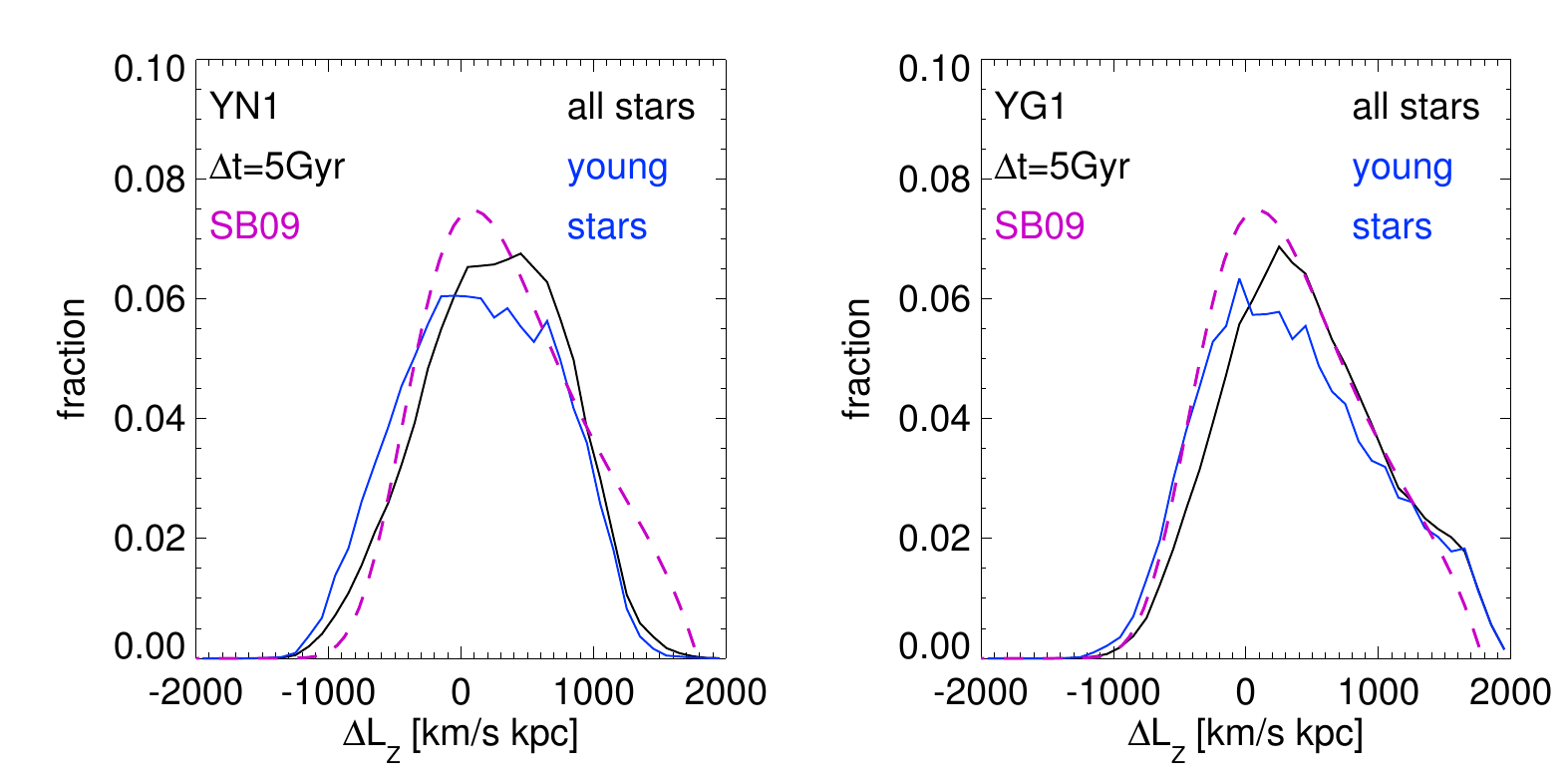}}
\caption{Plots of changes in $L_z$ in the last $5\Gyr$ for two samples of
stars selected to have $L_z=L_{z,{\rm circ}}(8\kpc)\pm100\kpc\kms$ 
at $t=t_{\rm f}$ in models YN1 (left) and YG1 (right). The black
curves are for all stars that are more than $5\Gyr$ old, while the blue
curves are for stars with ages $5-6\Gyr$, so $\Delta L_z$ is the change in
their angular momentum since they were young.  Also plotted in a broken line is the
corresponding distribution inferred by SB09a from the chemical
composition of Snhd.}\label{fig:DLz}
\end{figure*}

\subsection{Vertical dispersions profiles over time}
\label{vertdp}

We have already shown that most models with GMCs show exponential
vertical profiles which lack thick components. Among our Y models, we find
$\sim$ five models which have noticeable wings at $|z|\gtrsim1\kpc$ in their
vertical density profiles, suggestive of a thick disc.  Apart from Y3f, which
has a very mild wing, all these models share with Y4 a growth history in
which the scalelength for star insertion rises only to $h_{R,\rm f}=2.2\kpc$,
so the disc is fed compactly. Model Y4f$\zeta-$ stands out amongst these
models in having a vertical profile that rather nearly matches that of our
Galaxy up to $z=2\kpc$, but still falls off much more steeply at higher
altitudes.  Apart from a compactly fed disc, Y4f$\zeta-$ has a fixed inner cutoff
and a star formation efficiency of only 4 per cent, so it combines strong GMC
heating with strong spiral structure due to the compact feeding.

In this section, we compare Y4f$\zeta-$ to one of the standard GMC models,
Y1, which has inside-out growth to $h_{R,\rm f}=4.3\kpc$, star-formation
efficiency $\zeta=0.08$ and an adaptive cutoff, to understand the origin of
the different vertical profiles. In the right hand panels of
Fig.~\ref{fig:rhozt} we plot the vertical component of velocity dispersion
for stars of all ages at $R=8\kpc$ in both models.  In Y1 $\sigma_z$ grows
steadily, while in Y4f$\zeta-$ steady growth of $\sigma_z$ is interrupted by
a jump at $t\sim5\Gyr$.

The left hand panels of Fig.~\ref{fig:rhozt} show the vertical density profiles
of these models at $R=8\kpc$ at several times between $1$ (black) and
$10\Gyr$ (yellow). Notwithstanding the steady growth in $\sigma_z$, the
vertical profile of Y1 changes remarkably little. This is in part because
heating by GMCs is offset by adiabatic contraction as the disc gains mass,
and in part because the mid-plane density is being constantly increased by
the addition of star particles. The sudden increase in $\sigma_z$ at
$t\sim5.5\Gyr$ \emph{does} cause the vertical profile of Y4f$\zeta-$ to change
significantly, because the sudden increase in $\sigma_z$ is not offset by a
sudden increase in surface density or enhancement of density at $z=0$.

The sudden increase in $\sigma_z$ is caused by a strong and extended $m=3$
mode developing in its disc, followed by a large bar structure. These
non-axisymmetric structures cause strong radial redistribution of stars and
lead to final scalelengths of the disc that are significantly larger than the
compact input scalelengths.  We find that using a fixed cutoff for Y4
models leads to more extended non-axisymmetric structures and thus more
strongly thickened discs, than an adaptive cutoff. For the Y models with
milder thick wings we find that their thick components arise from similar
processes of smaller magnitude.

\subsection{Radial migration}\label{sec:Rmigrate}

When a star is scattered by a non-axisymmetric structure across the
structure's corotation resonance, its angular momentum changes appreciably
without any increase in the eccentricity of its orbit
\citep{sellwoodb}. Since a change in $L_z$ corresponds to a change in the
star's guiding-centre radius $R_{\rm g}$, this process drives radial migration.
Fig.~\ref{fig:DLz} shows for two models, YN1 and YG1, histograms of the change
$\Delta L_z\equiv L_z(10\Gyr)-L_z(5\Gyr)$ in the angular momentum of stars
which at $t=t_{\rm f}$ have angular momentum in the range $L_z=L_{z,{\rm circ}}(8\kpc)\pm100\kpc\kms$.
Here $L_{z,{\rm circ}}\equiv R v_{\rm circ}(R)$ and these stars thus have $R_{\rm g}\approx8\kpc$.
The black curves include all stars older than $5\Gyr$, while the blue curves are for
stars with ages $5-6\Gyr$, so $\Delta L_z$ is the change in
their angular momentum since they were young.

\citet{minchev10} claim that bars can contribute to radial migration.
Fig.~\ref{fig:DLz}, however, reveals that the histogram of $\Delta L_z$
values for a model, YN1, with one of the strongest bars, shows a sharp drop
at high positive $\Delta L_z$. For young stars this cutoff simply reflects
the fact all such stars have started from near-circular orbits at $R>R_{\rm
cut}(5\Gyr)=3.3\kpc$, so stars now near $R_{\rm g}=8\kpc$ have $\Delta L_z\la
v_c\times(8-3.3)\kpc$. Older stars could have reached $R_{\rm g}=8\kpc$ from
any radius in the disc, so the fact that their histogram of $\Delta L_z$
values shows the same cutoff implies that stars are not escaping from within
the bar (which by construction extends to $R\sim R_{\rm cut}$). This
conclusion is reinforced by the fact that a model such as YG1, which develops
a short bar very late on, has a histogram of $\Delta L_z$ values that has a
peak of comparable width to that of YN1, plus an extension at fraction
$\sim0.02$ that reaches to values of $\Delta L_z$ that are $\sim500\kpc\kms$
lager than those reached by the histogram for YN1. We conclude that in our
simulations bars play a minor role for radial migration to $R=8\kpc$.

\begin{figure*}
\centerline{\includegraphics[width=14cm]{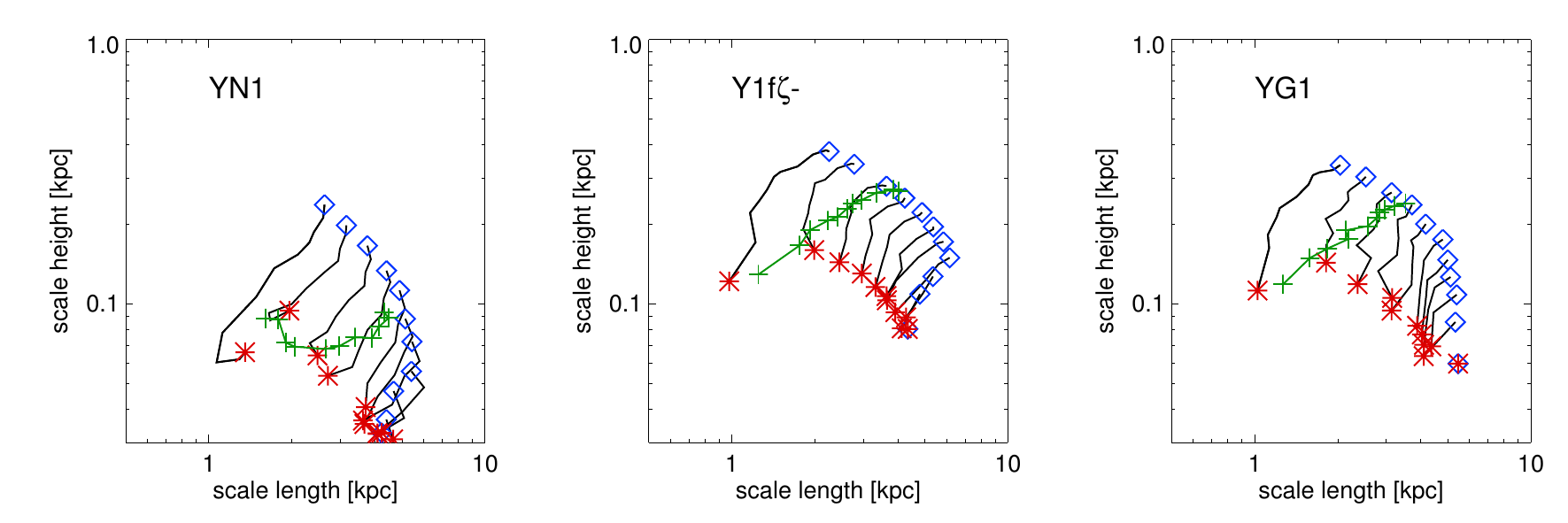}}
\caption{Each black curve shows fitted values of $h_R$ and $h_z$ to the
spatial distribution of an almost coeval population at a series of times: 
a red star marks values at the earliest time (when ages lie in $(0,1)\Gyr$)
and a blue diamond marks values during the last Gyr of the simulation. 
For $h_R$ we consider the region $R=6-10\kpc$ and for $h_z$ we use
the vertical profile at $R=(8 \pm 1) \kpc$. In every case the particles of the 
IC yield the curve at top left. Fits to the distribution of all stars, 
regardless of age, are marked by green crosses.}\label{fig:sgrow}
\end{figure*}

SB09a showed that radial migration enables us to understand the chemical 
composition and widening age-metallicity relation of the Snhd in a natural
way from basic chemical evolution: Metal advection by the inwards directed
galactic flow and (to a lesser degree) the faster and earlier star formation
make the inner regions of the galaxy significantly more metal rich than the
outskirts. Radial migration brings those stars now to the solar neighbourhood
on timescales of a few Gyrs, while those locally measured stars have only 
small differences in mean asymmetric drift vs. metallicity. 
It should be noted that for the models shown in Fig.~\ref{fig:DLz} 
$L_{z,{\rm circ}}(R=8\kpc)\sim1900\kms\kpc$, which means that the stars with 
the highest $\Delta L_z$ in model YG1 had very small $L_z$ at $t=5\Gyr$. So,
radial migration can bring stars from the inner regions of a galaxy to the outer disc
on cosmological timescales. This has been indirectly inferred from observations
of stars in the Snhd that are more metal rich than the Sun but old and
yet not on highly eccentric orbits \citep{casagrande,kordopatis} .

SB09a used their chemodynamical evolution model to determine the required
strength of radial migration. A parameter was used to control the width of
the $\Delta L_z$ distribution, and by fitting their model to the local
chemistry they could determine the optimum value of this parameter and thus
predict the $\Delta L_z$ distribution. The magenta dashed curves in
Fig.~\ref{fig:DLz} show this prediction. The agreement between this predicted
distribution and that measured in model YG1 (and many other models) is
remarkable considering the different physical principles generating each
distribution.  In fact, the differences between the SB09a predictions and the
models are smaller than the uncertainties for these comparisons arising from
differences in e.g. radial density distributions or rotation curves --
assumed in SB09a,b to be flat at $v_c=220\kms$.

The overwhelming majority of our Y models yield histograms of $\Delta L_z$
that are similar to one of those shown in Fig.~\ref{fig:DLz} or are
intermediate between these two cases. Models which have strong bars
already in place at $t=5\Gyr$ yield $\Delta L_z$ distributions like that of YN1. These
include models without GMCs, such as YN1 and YN3 and models with $\zeta=0.16$
and thus lower total GMC mass, such as Y1f$\zeta +$ or Y1f$\zeta +$m2. In the
majority of Y models with GMCs bars are weaker and form at later times (see
also Section \ref{seceff}), but they show well developed spiral structure and
their distributions of $\Delta L_z$ are more similar to that of YG1.  Hence
strong deviations from the predicted $\Delta L_z$ distribution of SB09a are rare. An
unusually narrow histogram is produced by model Y2Mb-, which has an
anomalously low-mass disc ($M_{\rm f}=3\times10^{10}\msun$) and lacks
inside-out formation ($h_{R,\rm i}=h_{R,\rm f}=2.5\kpc$). Models Y2Mb+, which
has a higher disc mass and Y4f$\zeta -$, which develops extended
non-axisymmetric structures, show mildly wider $\Delta L_z$ distributions.

In non-standard models with halo F (see Section \ref{lowhalo}), very extended
$m=2$ structures can develop which can lead to noticeable excesses of stars
at $\Delta L_z<1000\kms\kpc$ (e.g. models FN1, F2l). For models with thick 
disc ICs and GMCs (see Section \ref{sec:thkD}), strong bars tend to form 
earlier than in Y models and thus a higher fraction of models, including
e.g. Z1 and A1$\tau$, show a sharp drop at high positive $\Delta L_z$ similar to YN1.
A detailed  discussion of the dependence of radial migration on the
disc's evolution history is a topic for a future paper.

In contrast to the models of SB09b, where old, kinematically hot stars form a
thick disc, our simulations do not develop thick discs. This arises from the
modelling of the vertical heating applied in SB09b, which differs from our
simulation in three respects. (i) The assumption of vertical energy
conservation in SB09a was disproven by \citet{solway}, who demonstrated that
action rather than energy was conserved. (ii) SB09a used for each population
a heating law that depends only on birth radius, while stars are heated
throughout their trajectory through the galaxy. With a given heating law,
this assumption exaggerates the difference between the final random
velocities of stars born at small and large radii.  (iii) SB09a assumed a
fixed radial dependence of the heating to ensure constant scale height of the
local populations.  Preliminary examinations of our simulations however show
less inner disc heating and also a different time dependence.  A detailed
discussion of heating rates will follow in a subsequent paper.

\subsection{Scalelength growth}\label{sec:Rhgrowth}

Each black curve in a panel of Fig.~\ref{fig:sgrow} shows the exponential scalelengths
$(h_R,h_z)$ recovered at different times for particles that were inserted in
a time interval $1\Gyr$ long. Particles present in the initial conditions
form the oldest cohort, and their curve lies at the upper left end of the
series of curves. The next oldest cohort is formed by particles added in the
simulation's first gigayear, etc. Somewhat analogously, \citet{bovyrix}
plotted a point in the $(h_R,h_z)$ plane for each ``monoabundance''
population of the Galaxy, and found that the points of ``$\alpha$-old''
populations tended to lie above and to the left of the points of
``$\alpha$-young'' populations. In the spirit of that study, we determine
$h_z$ by fitting single exponentials to the distribution in $z$ of particles
that lie in the cylindrical shell $2\kpc$ wide around $R=8\kpc$. We determine
$h_R$ by fitting single exponentials to the radial surface density profile  
of particles that lie at $R=6-10\kpc$. Each cohort's red star shows
$(h_R,h_z)$ just after the birth of the cohort, and the blue diamond shows
the values at the end of the simulation. Hence the blue diamonds are what
one might compare with the points of \citet{bovyrix}.

The horizontal location of the red stars simply reflects the rule used to add
particles. When a model has inside-out growth (e.g., Y1f$\zeta -$) the red stars march
to the right, but in other cases (e.g., Y2) they do not.  The vertical
locations of red stars reflect the velocity dispersions of young stars, and
they tend to move downwards over time because as the disc gains surface
density, the velocity dispersion that can be acquired in $\sim0.5\Gyr$ allows
particles to move less far from the plane. For nearly every cohort in nearly
every model, the scalelengths $(h_R,h_z)$ at birth are significantly smaller
than their values at the end of the simulation. Exceptions to this rule are
models such as Y2Mb- that have anomalously low-mass discs and consequently
develop only weak spiral structure, so that $h_z$ increases at roughly constant
$h_R$.

\begin{figure}
\centerline{
\includegraphics[width=8cm]{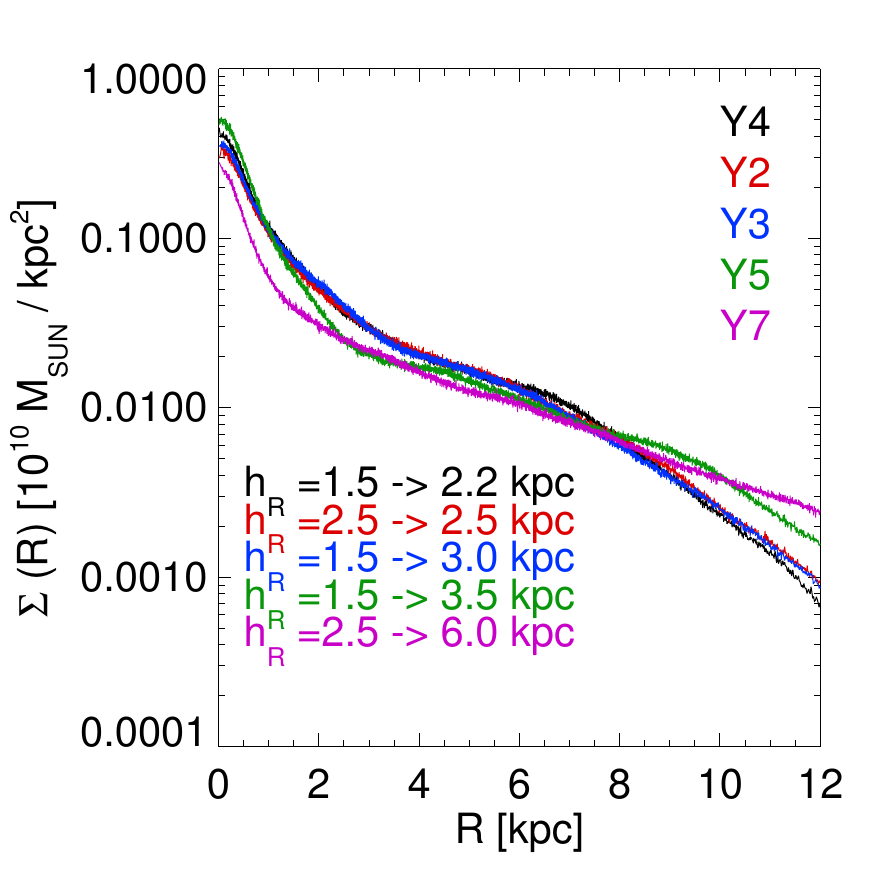}}
\caption{Face-on stellar surface density profiles $\Sigma(R)$ for models, which differ only
in their radial growth history $h_R(t)$. They all grow with time as $t^{0.5}$.}
\label{figX}
\end{figure}

In Fig.~\ref{fig:sgrow} the leftmost panel for YN1 is notable for the small
values of $h_z$ that arise because this simulation has no GMCs. The middle
panel for Y1f$\zeta -$ has more realistic values of $h_z$ because it has GMCs. The
panel on the extreme right, for YG1, shows somewhat less growth in $h_R$
because the introduction of gas has weakened the bar.

The green crosses in Fig.~\ref{fig:sgrow} show the values of $(h_R,h_z)$
obtained by fitting all stars present at each epoch, regardless of their age.
When there are no GMCs, $h_R$ grows but $h_z$ does not (left panel), while
both scalelengths grow in concert when GMCs are present.

In Fig.~\ref{figX}, we show the resulting surface density profiles from five
models, which only differ in their radial growth history $h_R(t)$. They all
grow with time as $t^{\xi=0.5}$.  We find that the three compact models Y2,
Y3 and Y4 have almost indistinguishable radial profiles. The more extended
models Y5 and Y7 have smaller central regions of enhanced surface density,
reflecting shorter bars.  In these models the bars are shorter because
surface densities are lower.  At $R\sim3-7\kpc$ all profiles are almost
parallel, i.e., have similar output scalelengths despite the different input
scalelengths. This radial redistribution is caused by non-axisymmetric disc
structures. The more compact models have down-turning breaks at $R\sim
8\kpc$, whereas Y5's profile turns down at $R\sim 10\kpc$ and Y7 is
exponential out to $R>30\kpc$.

Model Y1 is not shown because it is intermediate between Y5 and Y7.  Model
Y6 has a different time dependence of radial growth ($\xi=0.2$), but
otherwise the same parameters as Y3, and it hardly differs from Y3. In Y1f, Y3f
and Y4f the use of fixed rather than adaptive cutoffs makes the zones of
equal scalelengths broader ($R\sim3-9\kpc$) and the radial breaks in the profiles
of Y3f and Y4f move outwards to $R\sim9-10\kpc$.

\begin{figure*}
\centerline{
\includegraphics[width=14cm, angle=90]{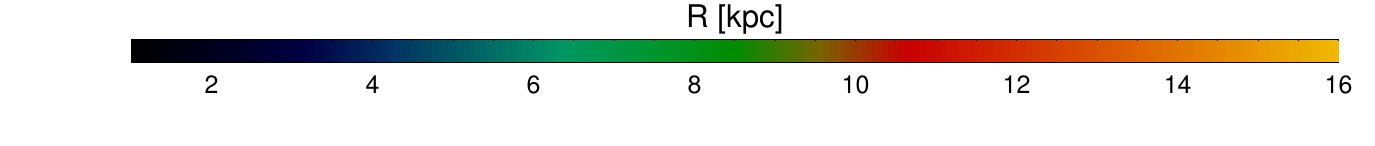}
\includegraphics[width=7cm]{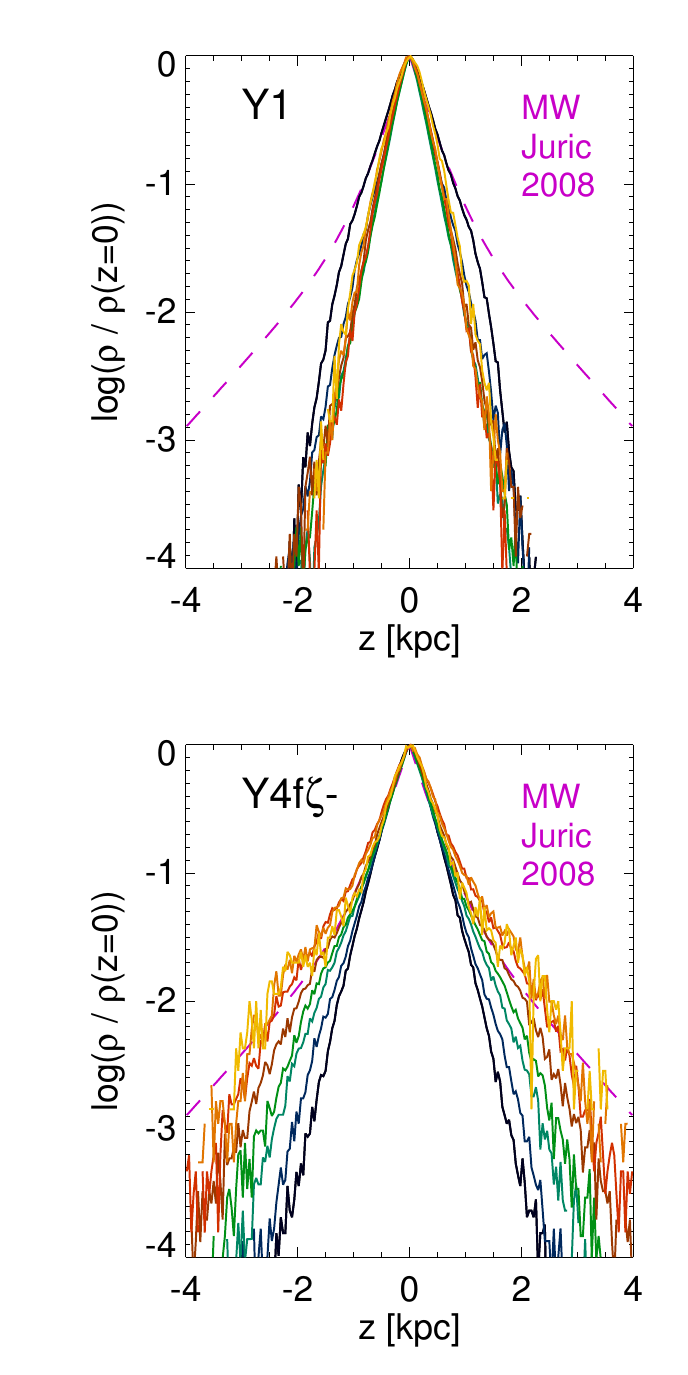}
\includegraphics[width=14cm, angle=90]{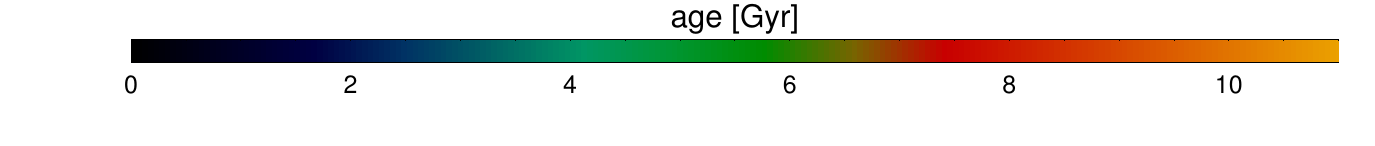}
\includegraphics[width=7cm]{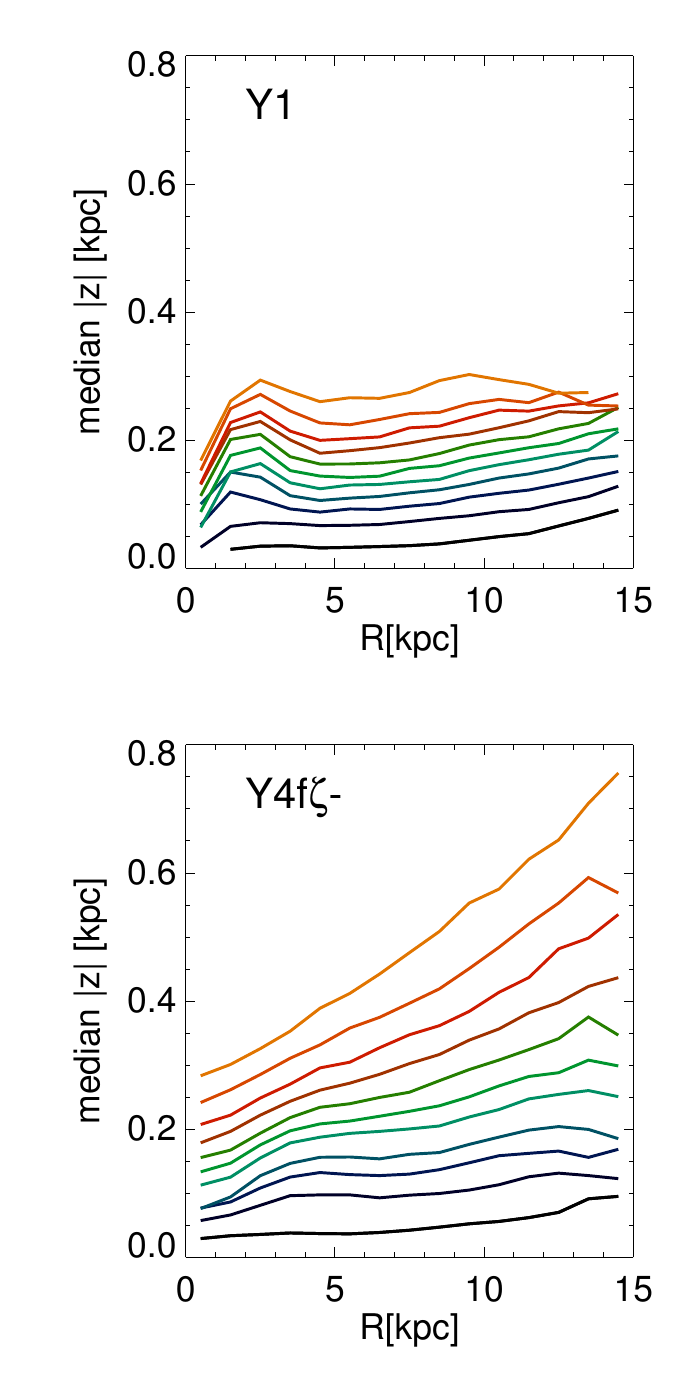}}
\caption{Left hand panels: radial dependence of the vertical profiles of models
Y1 and Y4f$\zeta -$ between $R=2$ (black) and $16\kpc$ (dark yellow) as
indicated on the colour bar at extreme left.
Right hand panels: The median altitude over the midplane $|z|$
of mono-age populations in models Y1 and Y4f$\zeta -$ as
a function of radius $R$. Here colour encodes age between 0 (black) and
11 Gyr (dark yellow) as indicated on the central colour bar.
}\label{fig:flare}
\end{figure*}

\subsection{Flaring of discs}

It is generally accepted that the vertical profiles of disc galaxies 
are very constant radially \citep{vdkruit}. The left hand panels of 
Fig.~\ref{fig:flare} examine this in the case for our final models. Lines
of various colours show profiles from $R=2\kpc$ (black) 
to $R=16\kpc$ (dark yellow).

Model Y1 has a buckled bar, which causes a mildly thicker, non-exponential
profile in the bar region. In all other regions the vertical profiles are
exponential and almost indistinguishable. As described in Section
\ref{vertdp}, the majority of our Y models with GMCs have single exponential
vertical profiles. Model Y1 is typical of this class. As discussed
in Section \ref{seceff}, not all of these models have buckled bars. In models
without a buckled bar (e.g. Y2, Y1$\zeta -$), the vertical profile is constant 
throughout the whole radial range.

As discussed in Section \ref{vertdp}, a minority of our Y models have deviations
from single exponential vertical profiles in the form of thick wings. 
Model Y4f$\zeta -$ is the most extreme of these models and thus behaves 
very differently in Fig.~\ref{fig:flare}. At inner radii the profile is
single-exponential, as the bar has not buckled. The profile becomes
significantly thicker with increasing $R$, and thickness increases
continuously with radius. The thicker profiles are not double exponential,
but rather flatten at $\left|z\right|=1\kpc$ and then fall off steeply at
higher altitudes. Models with milder thick wings, such as Y4f, 
also show flaring, but the effect is considerably milder than for Y4f$\zeta -$.

\citet{minchev15} using cosmological simulations by \citet{a13} and
\citet{martig} recently claimed that all mono-age populations of stars should
have vertical profiles flaring with radius. In this picture, inside out
formation keeps the total vertical profile roughly constant with radius and the thick
disc stars at a given altitude become younger with increasing radius. In the
right hand panels of Fig.~\ref{fig:flare} we show how the median altitude
$\left|z\right|$ changes with radius $R$ for stars in age bins between 0
(black) and 11 Gyr (IC stars, dark yellow).

We notice that at all radii the median $\left|z\right|$ increases
monotonically with age for both models as a consequence of the monotonically
increasing AVRs. Only the oldest component of the outer disc of Y1 behaves
differently. This population was initially very compact and at 10 Gyr still
has a very low density in the outer disc. This brings along a peculiar
population of very eccentric orbits, which can reach the outer regions, but
have less than average vertical action.

The youngest stars always flare, which results from the constant input
vertical velocity dispersion and the reduced vertical force at lower surface
densities in the outer disc. In Y1 bar buckling causes all populations older
than 1 Gyr to be thicker in the inner disc. Outside the bar region, all
populations except the IC stars show mild flaring. In models without a
buckled bar, the median $\left|z\right|$ of these populations would decrease
monotonically towards $R=0$.

Model Y4f$\zeta -$, which has a thick vertical structure, shows strong
flaring for the old and intermediate age components: the median
$\left|z\right|$ increases by factors in excess of $2$.

\section{Non-standard models}
\label{nost}

We have seen that models with a generous supply of massive GMCs and a
standard disc and a standard dark halo yield quite realistic bars and thin
discs under a range of reasonable assumptions regarding how particles are
added to the disc.  However, we have been conspicuously unsuccessful in
generating a thick disc. A burning question is whether our failure to create
a realistic thick disc arises from an inappropriate parametrisation of disc
growth or a poor choice of parameters. And if our failure cannot be ascribed
to either of these causes, how were thick discs made?

\subsection{A thick disc from the initial conditions?}\label{sec:thkD}

\begin{figure}
\centerline{\includegraphics[width=7cm]{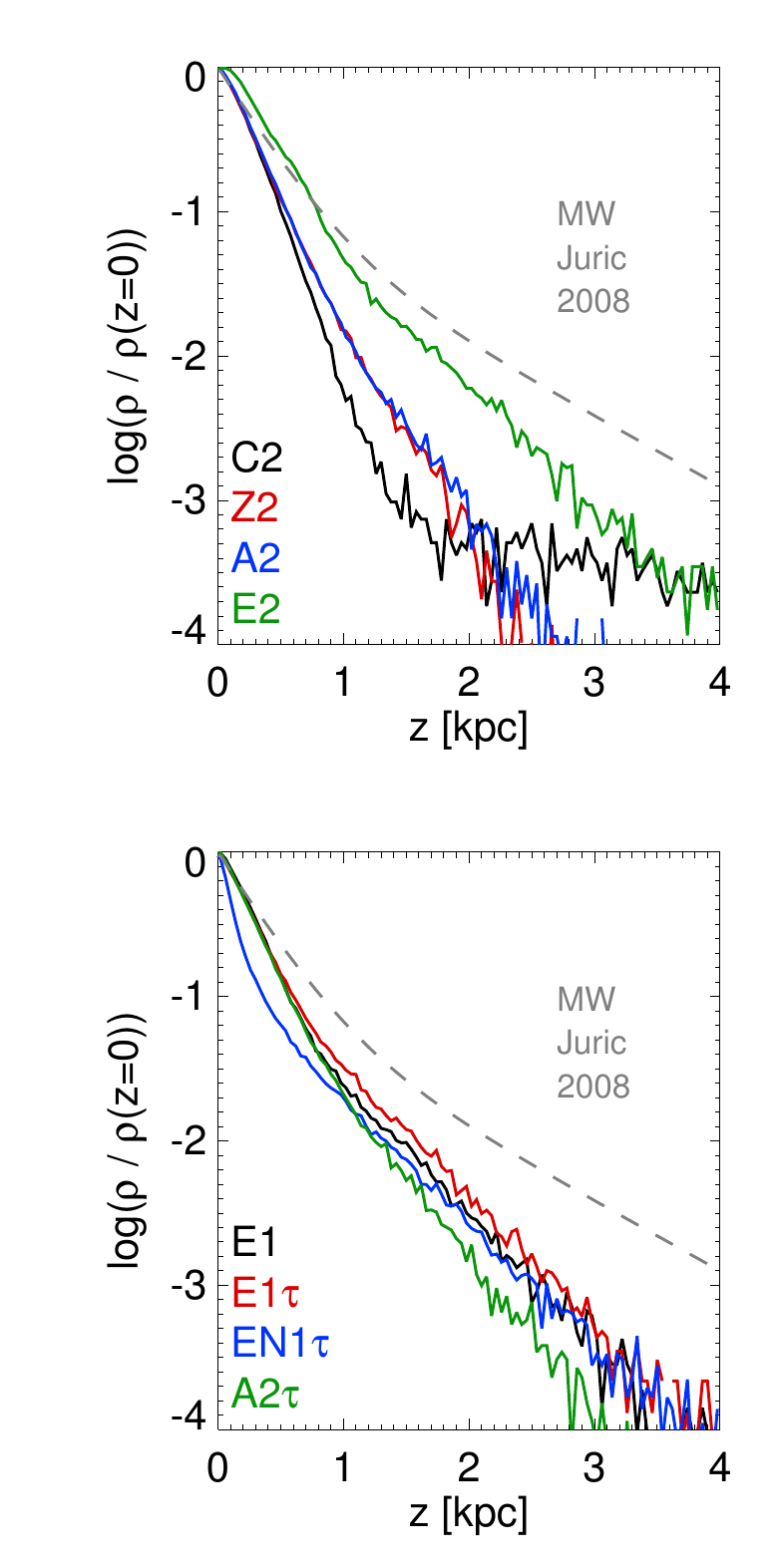}}
\caption{Vertical stellar density profiles at $R=8\kpc$ of models aimed at production
of a realistic thick disc. Top: results of adding a thick disc or bulge at
birth (models C2 - E2). Bottom: results of modifying the way the disc is fed
from the most successful IC in the top panel (E models) plus model A2$\tau$.
}\label{fig:thick}
\end{figure}
Since dynamics is not generating enough stars with large vertical actions, we
add such stars by hand. First we just add them to the ICs by including a
bulge (IC C) or a massive and extended thick disc (ICs Z, A, E). The upper
panel of Fig.~\ref{fig:thick} shows the vertical density profiles at
$R=8\kpc$ and illustrates the outcome of these experiments. The
only model to come near to the goal of generating an adequate thick disc is
E2, which starts from a disc with scalelengths $(h_{R,\rm disc},z_{0,\rm
disc})=(2.5,1.2)\kpc$ that is three times more massive ($M_{\rm
b,i}=1.5\times10^{10}$) than standard, so at birth we are endowing this
model with something very like the Galaxy's current thick disc. 

The essential challenge of matching the observed density of stars $1-2\kpc$
above the Sun is endowing the IC with enough stars with large values of the
vertical action $J_z$.  The IC disc is relatively low-mass, so the vertical
gravitational force it provides is not large, and stars with the required
values of $J_z$ must rise high above the midplane. Hence the structure
furnished by the ICs has to be fat. The structure also has to have a
significant extent radially because it does not grow much radially during the
formation of the thin disc -- as the mass inside a circular orbit grows, the
orbit's radius decreases.

In addition to not providing enough stars in the final model at
$|z|=2-4\kpc$, model E2 has a thin disc that is too thick, and a bar that is
too long ($\sim6\kpc$).  Therefore in the
lower panel of Fig.~\ref{fig:thick} we report the results of varying the way
particles are added to ICs A and E. These include (i) giving particles an
initial velocity dispersion $\sigma_0$ that declines from $36\kms$ to the
standard value $6\kms$ on a characteristic timescale $1.5\Gyr$ (models A2$\tau$,
E1$\tau$, EN1$\tau$), (ii) letting the scalelength for addition grow (from $h_{R,\rm
i}=2.5\kpc$ to $h_{R,\rm f}=4.3\kpc$ in E1 and E1$\tau$ and EN1$\tau$ rather than fixing it at
$2.5\kpc$ in E2), and (iii) excluding GMCs (EN1$\tau$). 

\begin{figure*}
\centerline{\includegraphics[width=16cm]{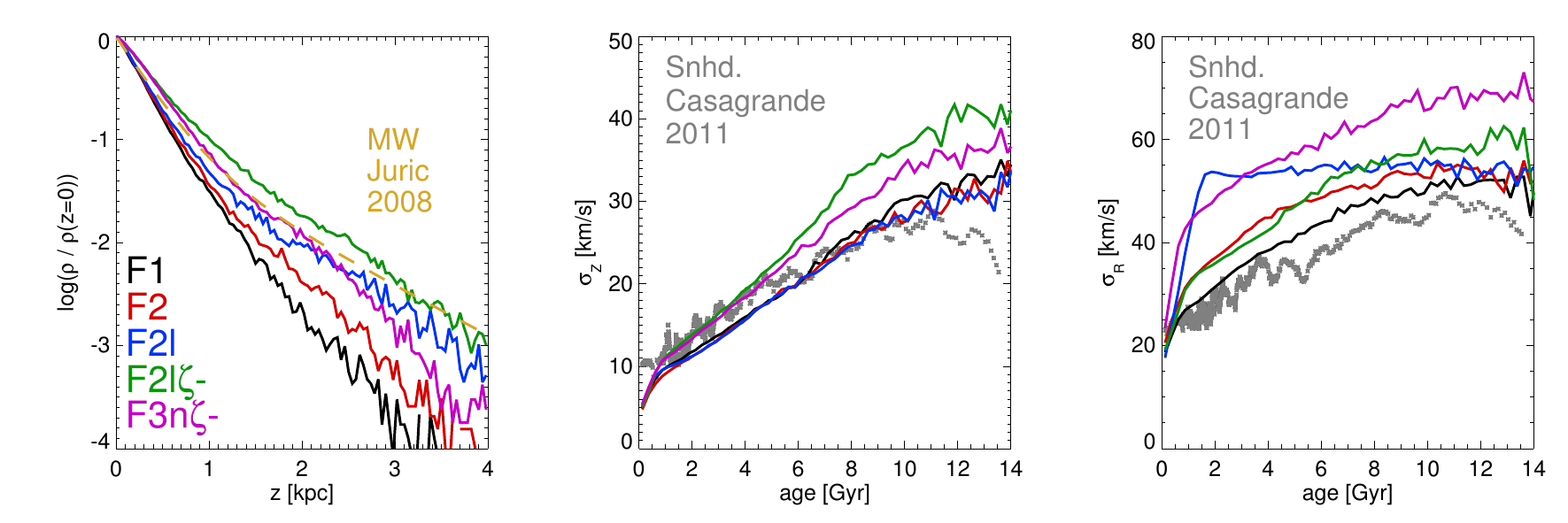}}
\caption{Characteristics at $R=8\kpc$ of models with the low-density halo F.
These models develop strong bars early on, and in consequence the
disc can become quite thick. Left panel: vertical density profiles. Centre
and right panels: $\sigma_z$ and $\sigma_r$ as functions of age after adjustment to
simulate the effects of observational uncertainties in the Snhd ages of
Casagrande et al.~(2011) shown as grey points. 
}\label{fig:lowD}
\end{figure*}

E1$\tau$ comes closest to the promise shown by E2, but again provides far too few stars at $|z|=1-4\kpc$.
It produces a thin disc in better agreement with the data as it has a shorter 
bar of length $\sim4.5\kpc$ that compares well with that in the MW. 
Assuming $\sigma_0\sim36\kms$ at early times in E1$\tau$ compared to always $\sigma_0=6\kms$
in E1 only mildly increases the number of stars at $|z|>1\kpc$. 
Model EN1$\tau$, which differs from E1$\tau$ in lacking GMCs, has a similar thick disc,
but, in common with other models without GMCs, a disc that is too thin.
A2$\tau$, which, like E1$\tau$ has $\sigma_0\sim36\kms$ at early times, is 
as bad at providing a thick disc as A2, which has $\sigma_0=6\kms$.
It follows that choosing $\sigma_0(t)$ as in A2$\tau$, E1$\tau$ and EN1$\tau$ has little effect
on creating a thick disc.

The vertical profiles in the E models are very constant with radius (with the 
previously discussed exception of buckled bars), as the thick discs are set in the ICs,
which by construction have radially constant vertical profiles, and the thin discs behave as in
a typical Y model.

From these experiments we conclude that the key to producing a thick disc is
to put in place already at $z=2$ a rather massive, extended and thick
component. By tweaking our parameters for the IC disc or applying even higher 
$\sigma_0(t)$ at early times, we judge that we could eventually
produce a thick disc that is compatible with the data. However, completion of
this programme would not in any way \emph{explain} the observed thick
disc. This situation contrasts with the situation as regards the thin disc
and the bar, which we feel are nicely explained by many of our standard
models.

\subsection{Varying halo density and disc mass}

\subsubsection{Low-density dark haloes}
\label{lowhalo}

We now consider the impact of using a cosmologically unorthodox dark halo. In
IC F the dark halo has an anomalously large scalelength $a_{\rm
halo}=51.7\kpc$ and in consequence a lower than standard density in the
visible galaxy (Fig.~\ref{fig:aHalo}). In IC G the dark halo has only half
the standard mass, so again in the star-forming galaxy the density of dark
matter is smaller than cosmology implies for the MW.  Reducing the halo's
density at fixed disc mass favours early formation of a powerful bar.  In
fact, model F2 produces a bar that extends to $R\sim10\kpc$. 

As in a standard halo, a model with strong inside-out growth (F1) develops a
vertical profile {at $R=8\kpc$} (Fig.~\ref{fig:lowD}), which deviates little from an exponential, while a more
compact model (F2) develops a stronger wing on its vertical density profile suggesting
a thick disc.  Adopting the AdapLi cutoff (described at the end of
Section~\ref{sec:addstar}) enhances the wing (F2l vs.\ F2).

The early and powerful bar, combined with a reduced contribution to the
vertical restoring force $K_z$ from the halo, makes the disc appreciably
thicker than the equivalent disc formed with a standard halo. In fact, some
models, for example F2l$\zeta-$, produce discs that are thicker than that of
the MW at $|z|<2\kpc$.  Unfortunately, in all our experiments a suitable thick disc, as in
model F3n$\zeta-$ out to $|z|=3\kpc$ (magenta lines in Fig.~\ref{fig:lowD}), which has no
cutoff and an enhanced supply of GMCs, is always associated with undesirable
features. One is a thin disc that is too thick.  Another is values of
$\sigma_z(\tau)$ at $R=8\kpc$ that are too large, especially for old ages. A
third is an over-long bar. On account of this long bar, $\sigma_R(\tau)$ at
$R=8\kpc$ is too large, especially at young ages.

The radial density profiles of the models with low-density haloes are diverse.
None resembles the distribution of added stars.  Counter-intuitively, the
AdapLi models (names containing 'l') are less concentrated than the standard
adaptive cutoff models: AdapLi models acquire high central surface densities
and powerful $m=2$ modes early-on; these modes then efficiently redistribute
stars in radius. 

As the amplitude $A_2$ quickly becomes strong in all F models, our procedure
for adding stars on nearly circular orbits is questionable in these models.
The particles do settle to orbits supported by the non-axisymmetric
potential, but the large value of $A_2$ significantly distorts the radial
distribution of recently introduced particles from that desired.  However,
large bars and strong low-$m$ spirals are an inevitable consequence of a high
baryon fraction \citep{deba, selcar}.

Similar to model Y4f$\zeta -$ shown in Fig.~\ref{fig:flare}, the models in 
halo F all have vertical profiles that are considerably thicker in the outer
regions of the discs compared to the inner regions. As none of the bars in 
these models buckles, the inner regions are always the thinnest parts of the models.

\subsubsection{Non-standard disc masses}

Finally we briefly consider the effects of increasing or decreasing $M_{\rm f}$,
the total baryon mass. Our standard mass $M_{\rm f}=5.0\times10^{10}\msun$ is in 
agreement with the $M_{\rm MW}=(5.6\pm1.6)\times10^{10}\msun$ determined by
\citet{piffl} for the MW.

\begin{figure}
\centerline{\includegraphics[width=7cm]{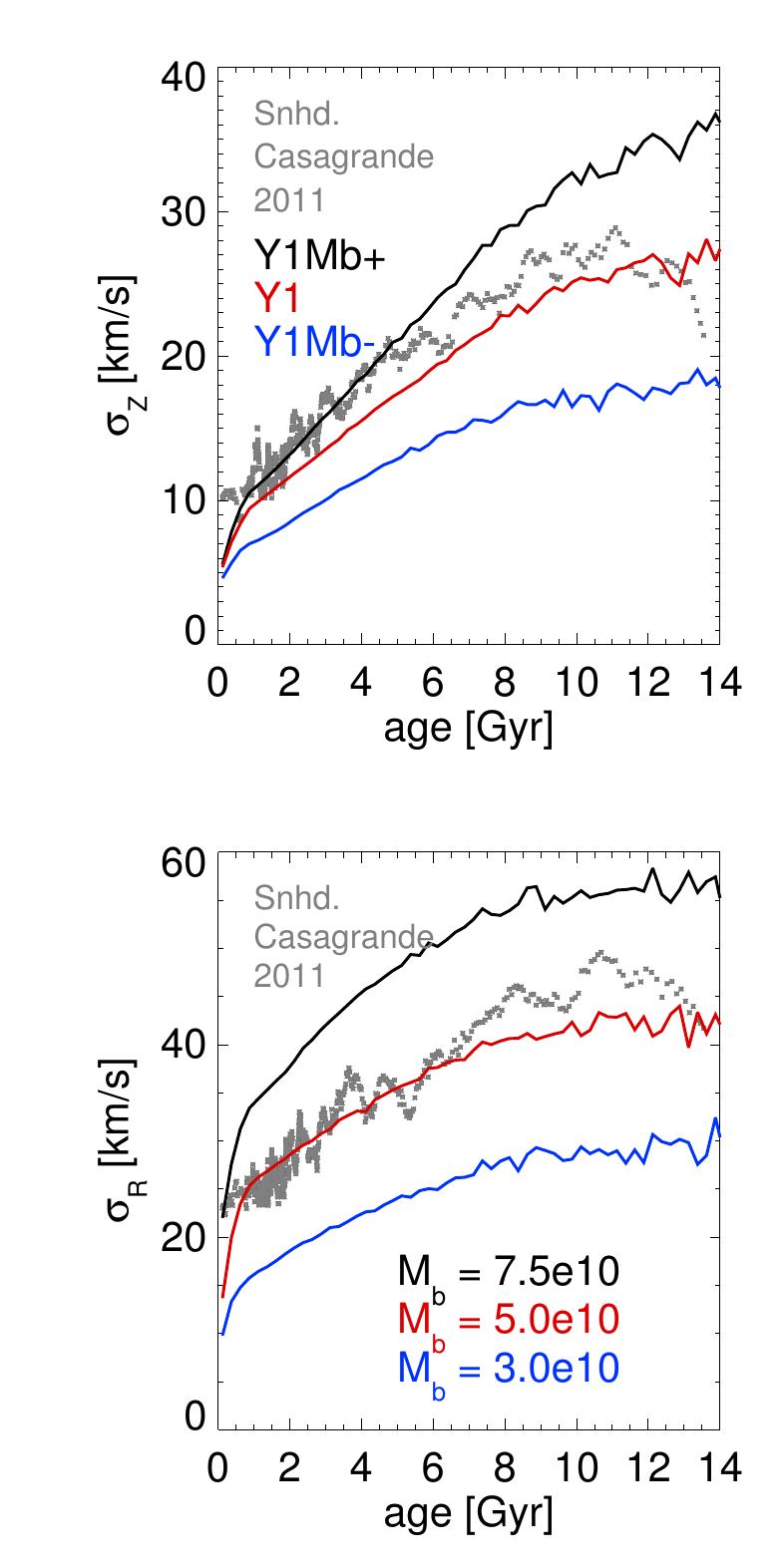}}
\caption{Velocity dispersion versus age at $R=8\kpc$ in models Y1Mb+, Y1 and Y1Mb-
which differ in total final disc mass. The data have been adjusted to
simulate the uncertainties and biases in the ages of Casagrande et al.~(2011)}\label{fig:Y3032}
\end{figure}

Models Y1Mb+ and Y2Mb+ have $M_{\rm f}$ 50 per cent greater than standard, while Y1Mb-
and Y2Mb- have $M_{\rm f}$ 40 per cent lower than standard. Increasing $M_{\rm f}$
strengthens non-axisymmetries, and vice versa when $M_{\rm f}$ is reduced. In
Section~\ref{sec:Rmigrate} we already noted that Y2Mb- has an unusually narrow
distribution of $\Delta L_z$ and in Section~\ref{sec:Rhgrowth} we noted that
in Y2Mb- shows an anomalously small increase in the radial scale length $h_R$
with which coeval particles are distributed between addition and the final
model. Both these results are immediate consequences of smaller departures
from axisymmetry. The deleterious effect of moving from the standard value of
$M_{\rm f}$ are nicely illustrated by Fig.~\ref{fig:Y3032}: non-axisymmetries
cause $\sigma_R(\tau)$ at $R=8\kpc$ to be too large at all ages in Y1Mb+ and too small at
all ages in Y1Mb-. Since GMCs couple horizontal motions to vertical motions,
$\sigma_z(\tau)$ at $R=8\kpc$ is always too small in Y1Mb-. In Y1Mb+ it is too large at old
ages. With the standard disc mass both $\sigma_R(\tau)$ and $\sigma_z(\tau)$
are about right.

It is worth noting that the scale-height $h_z$ of the disc is almost
independent of $M_{\rm f}$ because increasing $M_{\rm f}$ increases both
$\sigma_z$ and the surface density of the disc and thus the gravitational
force $K_z$ that together with $\sigma_z$ determines $h_z$.

These models show that our standard choices of disc and halo mass provide a
balance between disc and halo gravity that is close to that present in the
MW. Consequently, models Y1Mb+ and Y2Mb+ also have circular speeds at $R=8\kpc$
that are too high. To get the rotation-curve constraints right, one would
have to lower the halo density, which, as we have shown above, leads to
problems with overly long bars. For our standard mass values, 50 per cent of
the radial force at $R=8\kpc$ already comes from the dark halo.
Consequently, models Y1Mb- and Y2Mb- with lower disc masses have circular speeds
that are only marginally too low. If we assume that the GMC heating in models
Y1Mb+, Y1 and Y1Mb- is appropriate, the AVRs in Fig.~\ref{fig:Y3032} would
suggest the actual MW disc mass is slightly above $5\times10^{10}\msun$ and
the radial acceleration by the halo would have to be lowered mildly to
account for the additional disc mass. Due to the many uncertainties involved,
such as halo shape and disc mass profile, detailed estimates are not
appropriate here.

\section{Discussion}

Inevitably, our models of growing disc galaxies in live dark haloes
rely on idealised assumptions. The same is true for some of the analysis presented.
In this section we discuss the impact of several of these assumptions. 

For our analysis we have used azimuthal averaging to determine mass profiles
and velocity dispersions. Bars and spirals are non axisymmetric and also
defining features of galactic discs, both in observed galaxies and in our
models.  In the presence of a bar, stars will stream outwards at some
azimuths, and inwards at others. We have computed $\sigma_R$ as the standard
deviation of $v_R$ for all stars in a cylindrical shell, so any systematic
radial streaming contributes to our reported values of $\sigma_R$. We hope in
the near future to study the impact of the bar and spiral structure on the
kinematics of discs. In such a study, the distribution of values of $v_R$ at
any point would be decomposed into its mean $\overline{v}_R$ and dispersion,
and the latter would be smaller than the values $\sigma_R$ reported here by
$\sim\overline{v}_R^2/2\sigma_R$, which would be small for the anticipated
values $\overline{v}_R\la7\kms$. 

On account of non-axisymmetric structures, perfect circular orbits do not exist in disc galaxies.
However, we use the concept of an azimuthally averaged circular speed to insert
star particles into our disc galaxies on near-circular orbits. Outside the bar region
this is justified by the finding that the radial velocity dispersion of young stars
agrees with measurements in the Snhd, which is determined not by input $\sigma_0$ but 
by the local strength of non-axisymmetries. Within the bar region, we avoid inserting
new stars by introducing an inner cutoff $R_{\rm cut}$. This may seem a crude
assumption, but bar regions in disc galaxies are known to be deficient in
star formation \citep{sheth02}. There is plenty of SF in the central $200\pc$ of the MW (central 
molecular zone), but in our models we are not interested in these central areas
of disc galaxies, which are too small to affect the large scale dynamics studied here.
Moreover, models in halo Y which set $R_{\rm cut}=0$ do not yield results which 
differ significantly from those with $R_{\rm cut}>0$.

Our models deliberately lack a model for star formation from a gas component and for
gas circulation including accretion and feedback, as these processes bring along
both physical and numerical uncertainties of a severe nature. We find
that the specifics of the setup of the initial disc and the insertion of star
particles are unimportant for the outcome of our models, as long as the
inserted stellar populations are sufficiently cold. We have shown that the insertion
of GMCs has a much stronger effect on our models than the insertion of
isothermal gas. This is because gas merely assists stars in
forming spiral structure, which induces in-plane heating and radial
migration. Clouds, by contrast achieve something that neither stars nor gas
can: scatter stars out of the plane and thus increase the disc's vertical
scaleheight.

Although we have significantly updated the treatment of GMC populations
compared to previous studies by modelling a GMC mass function, applying
finite lifetimes, testing spatial clustering and having an evolving mass
fraction of GMCs, our GMC particles are clearly an idealised representation
of real GMCs. All our GMCs are assumed to have the same linear extent
regardless of their mass, whereas observed GMCs have sizes which on average
increase with mass, and the most massive GMCs can have radii in excess of the
$30\pc$ we adopt for the softening length of a GMC \citep{solomon}.  However,
real GMCs are not spherical and do not have spline-Kernel potentials, but are
strongly substructured with most of their mass concentrated in high-density
clumps \citep{blitz}.  As it is unfeasible to model this small-scale
structure in galaxy-wide simulations, we choose to make a simple assumption
about the potential of GMCs.

Another potential lack of realism is connected to the dark halo.  Although
our disc grows continuously, the mass of our halo is fixed.  Our Galaxy is
generally believed to have had a quiet merger history since $z>2$, and our
goal is to simulate evolution since the last major merger. The inner $10\kpc$
of dark haloes with such quiescent formation histories are expected to have
been almost completely in place by the end of the last major merger, as e.g.\
\citet{wang} have shown with high resolution cosmological simulations of
dark-matter-only haloes.  The dark-matter profiles in this region are
expected to be reshaped by baryonic processes (e.g.\ \citealp{pontzen}) as
they are in our simulations. In a forthcoming paper we plan to discuss the
impact on dark matter of growing discs and their bars and spirals.

Our models produce discs with single-exponential vertical profiles, whereas
the disc of the MW and comparable spiral galaxies have double-exponential
vertical profiles.  \citet{streich} recently presented observations of
low-mass disc galaxies with single exponential vertical profiles. Compared to
MW mass disc galaxies these objects are believed to have significantly lower
baryon fractions (e.g.  \citealp{moster}). As we have seen that thick discs
in our models are connected to higher baryon fractions, the question arises
whether our models in the standard halo Y on average have too low baryon
fractions. At late times there is little room to increase the mass as was
shown by the high mass models Y1Mb+ and Y2Mb+. At early times, our models have
relatively high fractions of the total baryonic mass (i.e. stars and GMCs)
in GMCs. The mass fraction is $\sim30$ per cent for a model with standard 
parameters such as Y1 and $\sim45$ per cent if $\zeta$ is lowered as in 
Y1$\zeta -$. Yet, these fractions are still smaller than some of the molecular
gas fractions reported at redshifts $z\sim2$ \citep{tacconi, scoville}.
Moreover, mass loss by stellar populations to the gas phase,
which is neglected in our models, is significant over $10\Gyr$, and thus old
populations at $10\Gyr$ could contribute higher mass fractions at early
times.  However, a significant part (depending on the IMF) occurs due to
core-collapse supernovae in the first $30\Myr$ of stellar evolution, which
for our purposes is negligible. Models with higher star formation rates
in the beginning (e.g. Y1fs3 vs. Y1f) have also been shown to not
significantly change our conclusions.

Our models have the same amount of radial migration as SB09a,b, but the stars
migrating outwards from the inner disc are not as hot vertically as in those
models, so our models do not produce thick discs.  This points towards the
necessity of an additional heating source for old/inner disc populations of
stars.  Several possible sources, both secular and external, have been
discussed in the literature: scattering of massive star forming clumps in
turbulent high redshift galaxies \citep{bournaud}, which can be regarded as
an extreme form of GMC heating, the continuous decrease of birth velocity
dispersion of stars in an interstellar medium which becomes less turbulent
over time \citep{forbes}, star formation during a phase of gas-rich merger
events \citep{brook} or the heating of an early thin disc by mergers
\citep{quinn}.

The idealised simulations of \citet{aw13}, in which galaxies formed from cooling 
gas inserted to substructured dark haloes from dark-matter-only cosmological
simulations, found that, in the absence of mergers, double exponential vertical
profiles can arise in rapidly evolving young galaxies, in which the central 
triaxial potential first has to be restructured into an axisymmetric disc-like 
potential, which can host near-circular orbits. The fully cosmological hydrodynamical
simulations of \citet{a13} showed that the periods of disc settling in which
the thin disc can survive over cosmological timescales, typically do not start
prior to $z\sim2$. The earlier periods are characterised by mergers, high
gas accretion rates and probably strong stellar feedback, which prevent
stable near-circular orbits. So, from a cosmological point of view, 
the existence of older and hotter stellar populations is certainly plausible.
As our experiments with thick disc ICs have shown, the appearance of these
objects at the start of thin disc settling would have been less disc like than
today.

\section{Conclusions}

We have presented a large set of controlled simulations of growing disc
galaxies within non-growing, live dark haloes. Most of these models start from
a thin, compact disc, with ten per cent of the final stellar mass, but we also
presented models with initially more or less massive discs, and thicker
discs. We also presented a model with a bulge rather than a disc in its
initial conditions.  The models grow by the constant addition of new star
particles on near-circular orbits in the midplane of the galaxy. At any time
particles are added according to an exponential radial density profile, but
the scale-length of this profile can grow over time. The majority of our
models contain a population of particles with masses in the range
$10^{5-7}\msun$ that represent giant molecular clouds (GMCs). These particles
are short lived and have masses drawn from a power-law mass function.  A
subset of our models contain an isothermal gas component to represent the
part of the interstellar medium that is not concentrated into GMCs.  Gas is
added to the isothermal component to hold roughly constant the fraction of
the disc's mass that is in smooth gas. 

We find that

\begin{itemize}

\item GMCs generate remarkably exponential vertical profiles. The scaleheight
of this exponential can match that measured for the thin disc of our Galaxy
providing the efficiency of star formation is at the lower end of the
estimated range and the masses of GMCs extend towards the upper end of the
range of estimated GMC masses. These exponential profiles are very constant
with radius, although buckled bars can produce deviations in the bar region.

\item Heating by GMCs is particularly effective early on, when the SFR is
high and the stellar disc is not yet massive. GMC heating significantly
delays, and in some cases can even prevent, the formation of a bar.

\item In order to suppress spurious two body heating from
      dark-matter particles to a negligible level, several millions of particles 
      in the live dark halo are needed. For such resolutions, GMC heating
      is significantly more effective and even in the absence of GMCs, 
      vertical profiles hardly change when particle numbers are increased.
      To resolve the vertical structures of young stellar populations, a
      force resolution at the sub-50 pc level is needed.

\item The role played by a smooth gas component is modest. It enhances the
efficiency of GMCs by increasing their effective masses. It also reduces the
lengths and increases the pattern speeds of bars, but only marginally.

\item Within most models, spiral structure drives a level of radial migration
within the disc that agrees well with estimates obtained by modelling the
chemical composition of the Snhd.

\item Unless the disc has an anomalously low mass, non-axisymmetric features
in the disc cause the final scale length of the disc, $h_R$, to exceed the
scalelength according to which particles have been added to the disc.

\item The disc's scaleheight $h_z$ soon settles to a value that does not
differ greatly from its final value. This is for two reasons: (i) the steady
increase in the vertical velocity dispersion of a cohort of coeval stars that
GMCs drive is partly offset by the disc's growing surface density, and self
gravity; (ii) freshly added stars are constantly reinforcing the density of
the disc near the plane.

\item Models in which the scalelength for mass insertion increases more
rapidly (stronger ``inside-out'' growth) have weaker non-axisymmetric structures
and on average shorter bars, and as a direct
consequence, smaller values of $\sigma_R/\sigma_z$ and thinner discs. By
contrast, the scalelength $h_R$ of the final disc is fairly insensitive to the
rapidity with which the scalelength for insertion grows, as longer bars cause more
radial redistribution.

\item The standard disc mass and halo parameters provide just the right level
of self-gravity in the disc. Increasing the mass of the disc yields values
of $\sigma_R$, $\sigma_z$ and $v_{\rm circ}$ at $R=8\kpc$ which are too high.
Reducing the mass of the disc leads to unacceptably low velocity dispersions 
and too little radial migration. Reducing the central density of the halo
leads to low $v_{\rm circ}$, to bars that arise too early and are too long, 
and to excessive values of $\sigma_R$ at $R=8\kpc$. 

\item In our models, a thick disc is very hard to form. The only dynamically 
generated thick discs we find generate the requisite vertical velocity dispersion
by O(1) departures from axisymmetry that extend to $R\sim10\kpc$.  In the case of the MW such
departures are ruled out by two facts: (i) O(1) departures generate in-plane
dispersions in relatively young stars that are too large, and (ii) they
generally make the thin disc too thick. Moreover, in these models
the vertical profiles are significantly thicker in the outer disc than in 
the inner disc.

\item We can obtain a structure that approaches the observed thin/thick
combination only by providing an essentially complete thick disc in the
initial conditions. The requirement that the primordial thick disc (PTD)
contain enough stars with the high values of the vertical action $J_z$ that
currently occur in the thick disc, causes the PTD to be so extended
vertically that it is not very disc-like. However, its radial extent must
also be considerable because stars of the PTD do not systematically increase
their guiding-centre radii: on average they do increase their angular momenta
$L_z$, but as the mass interior to an orbit increases, so does the value of
$L_z$ associated with a given guiding-centre radius.  Hence at $z=2$ the PTD
must already extend to $R\sim8\kpc$ given that the thick disc extends that
far now.

\end{itemize}

We are impressed that when we combine standard values for the disc mass and dark
halo parameters with observationally motivated assumptions about the
star-formation rate, the mass function of GMCs and the efficiency of star
formation, galaxies emerge through complex dynamics that bear a striking
resemblance to the MW: as functions of age, the horizontal and vertical
velocity dispersions at $R=8\kpc$ are similar to those observed, the
structure of the thin disc is about right, as are the length and strength of
the bar. Yet none of these quantities is prescribed by the admittedly rather
arbitrary manner in which we assemble the disc: changing the disc mass, the
halo density, the GMC mass function or the SFR efficiency from conventional
values destroys one aspect or another of the agreement between dynamical
model and observation. We consider that this outcome constitutes a
significant endorsement of $\Lambda$CDM cosmology and implies that our
prescriptions capture many of the essential features of the physics of galaxy
formation that shape disc galaxies. 

Our models provide a natural explanation of the observation that the
fraction of barred disc galaxies decreases with increasing redshift
\citep{sheth}: increasing mass renders a disc more vulnerable to bar
formation both because it puts the disc more in charge of the gravitational
field in which it rotates, and because the effectiveness of heating by GMCs
declines with the ratio of the mass of a GMC to the disc mass. In our
simulations the timing and violence of bar formation are critical for the
properties of the final disc. An issue for a subsequent paper is the impact
on the morphology of the bar of disc buckling after the bar has attained a
critical strength.

For our models, the only way to obtain a thick disc like that of  the MW appears to
be to add it to the ICs. This finding suggests that the MW's thick disc was
present already at $z\sim2$ and is a relic of the time before the MW settled
to its long period of secular growth.  It also strongly suggests that thick
disc formation requires additional sources of heat in addition to GMCs and
non-axisymmetric disc structures.  The additional heat may have been provided
by high-redshift mergers and/or it may have been inherited from strong
turbulent motions in the early disc. In contrast, GMCs and disc structure
apparently account for the full heating observed in the thin disc population
of the Snhd and thin components of external galaxies.

The MW is a more complex machine than can be adequately characterised by the
standard parameters, such as $h_R$, $h_{z,\rm thin}$, $h_{z,\rm thick}$,
$v_{\rm c}(R)$, $\sigma_R(R)$, etc. While not as complex as the MW, our
models are also too complex to be adequately characterised by a few
parameters. It will be interesting to compare mock observations drawn from
some of them to see whether they agree with the MW better than any naive
model, and to identify residual points of conflict. For example, the
circular-speed curve of the Galaxy is poorly known because the data are
significantly affected by spiral structure and the bar. It will be
interesting to  compare with data stellar kinematics and gas line-of-sight
velocities drawn from promising models. We hope to report on this exercise
shortly.

\section*{Acknowledgements}
We thank the referee for comments that helped improve the paper.
This work was supported by the UK Science and Technology Facilities Council (STFC) through grant ST/K00106X/1 
and by the European Research Council under the European Union's Seventh Framework Programme (FP7/2007-2013)/ERC grant agreement no.~321067.
This work used the following compute clusters of the STFC DiRAC HPC Facility (www.dirac.ac.uk):
i) The COSMA Data Centric system at Durham University, operated by the Institute for Computational Cosmology.
   This equipment was funded by a BIS National E-infrastructure capital grant ST/K00042X/1, STFC capital grant ST/K00087X/1, 
   DiRAC Operations grant ST/K003267/1 and Durham University. 
ii) The DiRAC Complexity system, operated by the University of Leicester IT Services. This equipment is funded by 
   BIS National E-Infrastructure capital grant ST/K000373/1 and STFC DiRAC Operations grant ST/K003259/1.
iii) The Oxford University Berg Cluster jointly funded by STFC, the Large Facilities Capital Fund of BIS and the University of Oxford. 
DiRAC is part of the National E-Infrastructure.

\label{lastpage}
\end{document}